\pdfoutput=1
\documentclass[epj]{svjour}
\usepackage{graphics}
\usepackage{microtype}
\usepackage[hyperindex=true,colorlinks=true]{hyperref}
\hypersetup{citecolor=[rgb]{0.125,0.679,0.196}} 
\usepackage{url}
\usepackage{doi}
\usepackage{slashed}
\usepackage{physics}
\newcommand{\spi}{\slashed{\pi}} 
\usepackage{feynmp-auto}
\usepackage{multirow}
\usepackage{enumitem}
\usepackage{amsmath}
\usepackage{amsfonts}
\usepackage{mathrsfs}
\newcommand{\exprime}{{\, \prime}}

\usepackage[numbers]{natbib}
\newcommand{\refeq}[1]{Eq.~\eqref{#1}}
\newcommand{\refeqs}[1]{Eqs.~\eqref{#1}}
\newcommand{\reftab}[1]{Table~\ref{#1}}
\newcommand{\reffig}[1]{Fig.~\ref{#1}}

\newcommand{\refsec}[1]{Sect.~\ref{#1}}

\newcommand{\bracket}[2]{{\langle #1 | #2 \rangle}}         
\newcommand{\bracketop}[3]{{\langle #1 | #2 | #3 \rangle}}  

\begin{document}
\title{Renormalization of pionless effective field theory in the $\mathrm{A}$-body sector}

\author{M. Drissi\inst{1}\fnmsep\thanks{\emph{Present address :} Department of Physics, University of Surrey, Guildford GU2 7XH, United Kingdom. \emph{Mail : }m.drissi@surrey.ac.uk}%
        \and T. Duguet\inst{1}\fnmsep\inst{2}
        \and V. Som\`a\inst{1}}                     
%
%
\institute{IRFU, CEA, Universit\'e Paris-Saclay, 91191 Gif-sur-Yvette, France
          \and KU Leuven, Institut voor Kern- en Stralingsfysica, 3001 Leuven, Belgium}
\date{Received: date / Revised version: date}
%
\abstract{
Current models of inter-nucleon interactions are built within the frame of Effective Field Theories (EFTs).
Contrary to traditional nuclear potentials, EFT interactions require a renormalization of their
parameters in order to derive meaningful estimations of observable. In this paper, a renormalization procedure is designed in connection with many-body approximations applicable to large-$\mathrm{A}$ systems and formulated within the frame of many-body perturbation theory. The procedure is shown to generate counterterms that are independent of the targeted $\mathrm{A}$-body sector. As an example, the procedure is applied to the random phase approximation. This work constitutes one step towards the design of a practical EFT for many-body systems.
} 
\maketitle
\section{Introduction}
\label{intro}

The problem of describing accurately and systematically nuclear systems from
their $\mathrm{A}$ constitutive protons and neutrons is now almost a century old. Starting with
the discovery of the neutron in the 1930s, the nuclear many-body problem remains a great challenge in spite of the development of a large portfolio of theoretical methods (see Ref.~\cite{Soma2018} for a recent historical recapitulation).

Almost 30 years ago, the seminal papers of Weinberg~\cite{Weinberg1990,Weinberg1991a}
and Rho~\cite{Rho1991} initiated a shift of paradigm. Rather than keeping on refining
interaction potentials between nucleons by fine-tuning their short-range behaviour,
the focus has shifted towards the development and the study of various effective
field theories (see Ref.~\cite{VanKolck2019} for a recent and pedagogical introduction to EFTs in the nuclear physics context and Ref.~\cite{Hammer2019EFT} for a recent review).
Within this theoretical framework, the computation of nuclear observables has to face the demand
of renormalizability.

In this article, the example of pionless effective field theory ($\spi$EFT)~\cite{Bedaque98} at leading order (LO) is used to demonstrate how the procedure
of renormalization of the nuclear Hamiltonian affects the calculations of many-body nuclear observables. One key aspect in this respect relates to the fact that, while the Schr\"odinger equation can be solved exactly in few-body systems (as required by the power counting at LO), it is and will remain practically impossible to do so in large-$\mathrm{A}$ sectors. This feature must be explicitly considered when applying EFT to arbitrary nuclear systems and requiring an order-by-order renormalizability.
In \refsec{sec:RenoPbMB}, the renormalizability issue and its consequences on $\mathrm{A}$-body calculations are introduced.
A renormalization tailored to a given many-body approximation is advocated as a crucial step toward the construction of EFTs for nuclear many-body systems at low energy. The problem is then addressed in a generic way for many-body approximations formulated within the frame of many-body perturbation theory (MBPT). In \refsec{sec:BasicTheo}, some essential theoretical tools, \textit{i.e.}\ Weinberg's asymptotic theorem~\cite{Weinberg1960} and Bogoliubov-Parasiuk-Hepp-Zimmermann's (BPHZ) theorem~\cite{Bogoliubov1957,Hepp1966,Hahn1968,Zimmermann1968,Zimmermann1969,Lowenstein1975} are briefly recalled. In \refsec{sec:RenoMB}, a general procedure to renormalize the Hamiltonian is derived for a given many-body approximation and independently of the $\mathrm{A}$-body sector of interest. In~\refsec{sec:AppAndExt}, the procedure is applied to the Random Phase Approximation (RPA). Eventually, extensions of this approach and the potential impact on the shape of future many-body approximations are discussed in \refsec{sec:Ccl}.

\section{Generalities}
\label{sec:RenoPbMB}

In this section, the problem of renormalization of the nuclear Hamiltonian
is introduced. First, emphasis is put on why it is becoming an important problem
to be explicitly addressed in many-body calculations. Second, the formalism to study
renormalization in a given many-body approximation is set up.

\subsection{The nuclear many-body problem}

Traditional many-body approaches rely on the given of a nuclear Hamiltonian $H$
and aim at computing its exact eigenstates $\ket{\Psi^\mathrm{A}_m}$
and their associated eigenvalues $E^\mathrm{A}_m$ in all $\mathrm{A}$-body sectors of interest,
\textit{i.e.}\ the goal is to solve the Schr\"{o}dinger equation
\begin{equation}
  H \ket{\Psi^\mathrm{A}_m} = E^\mathrm{A}_m \ket{\Psi^\mathrm{A}_m} \ \label{exactSE}
\end{equation}
where $m$ indexes the set of solutions, to the best accuracy possible.

In this context, the Hamiltonian can be modeled in various ways.
The current paradigm consists of building $H$ within the framework of chiral effective field theory
($\chi$EFT)~\cite{Weinberg1979,Weinberg1990,Rho1991,Weinberg1991a,Ordonez92} such that it takes the form of a series
\begin{equation}
  H_{\chi} \equiv H^{\text{LO}}_{\chi} + H^{\text{SL}O}_{\chi}
    = H^{\text{LO}}_{\chi} + \sum^{\infty}_{p=1} H^{\text{N$^{p}$LO}}_{\chi}\
\end{equation}
where the leading-order (LO) and the sub-leading orders (SLOs) are organised
according to a set of power-counting (PC) rules. First to be proposed historically,
Weinberg's power counting~\cite{Weinberg1990,Weinberg1991a,Ordonez92} happens to fit
traditional many-body calculations, \textit{i.e.}\ independently of the order at which
SLOs are truncated, \refeq{exactSE} is meant to be solved exactly to access
observables such as $E^\mathrm{A}_m$. However, Weinberg's PC has since been shown to violate the demand that the EFT is (order-by-order) renormalizable~\cite{Kaplan1996} and alternative PCs have been
proposed~\cite{Bedaque2002a}.  In addition to modifying the order at which certain
contributions enter the Hamiltonian, new PCs stipulate that, while LO
is to be solved exactly according to traditional many-body calculations,
SLOs must be computed in perturbation with respect to the LO solution.
It happens that the same computational scheme underlies the PC at play in $\spi$EFT~\cite{Hammer2019EFT}.

Generally speaking, the order-by-order renormalizability means that, at each order, observables can display a dependence on the regularization, \textit{e.g.}\ in the form of a dependence on the ultraviolet cut-off, that is not larger than the intrinsic uncertainty carried at the working order. To achieve that, observables are enforced to depend, at each order, on inverse powers of the cut-off and reach a finite limit as the cut-off is sent to infinity. In this context, $\spi$EFT has been shown analytically to satisfy renormalizability at LO up to three-body systems~\cite{Bedaque2000} and numerically in the case of four-body systems~\cite{Platter2004}.

\subsection{Renormalization and many-body approximations}

To this day, solving exactly \refeq{exactSE} for $H^{\text{LO}}_{\spi}$ remains in general
numerically intractable for $\mathrm{A} \gg 10$. Consequently, one \emph{must} design an additional expansion and truncation when
attempting to solve
\begin{equation}
  H^{\text{LO}}_{\spi} \ket{\Psi^\mathrm{A}_m}^{\text{(LO)}}
    = E^{\mathrm{A}\text{(LO)}}_m \ket{\Psi^\mathrm{A}_m}^{\text{(LO)}} \ . \label{exactLOSE}
\end{equation}
Typical truncations applicable to $\mathrm{A}$-body systems with $\mathrm{A} \gg 10$
are nowadays implemented on the basis of non-perturbative self-consistent Green's function
(SCGF)~\cite{Dickhoff04,Carbone13,Soma14b}, coupled cluster (CC)~\cite{Hagen14,Binder14}
and in-medium similarity renormalization group (IM-SRG)~\cite{Tsukiyama2011,hergert16a} methods
but also on the basis of MBPT~\cite{Goldstone1957,Ti16,Tichai:2018mll}.

Traditionally, and in agreement with power-counting rules of $\spi$EFT, $H^{\text{LO}}_{\spi}$ is renormalized on the basis of an all-order calculation in two- and three-body sectors. This happens to be technically feasible. Given, however, the impossibility to generate exact calculations in large $\mathrm{A}$ sectors, investigating how observables are impacted by the mandatory approximations appears to be a necessary task to validate the practical use of the current form of $\spi$EFT across a large fraction of the nuclear chart. More specifically, one must question to which extent the approximate solving of \refeq{exactLOSE}, on the basis of a previously renormalized potential via an exact calculation in two- and three-body sectors, compromises renormalization invariance. In the next step, the rationale must be further extended to SLOs in agreement with the PC at play.

In this article, the problem is attacked via reverse engineering, \textit{i.e.}\ instead of checking whether the renormalization invariance of observables obtained from $H^{\text{LO}}_{\spi}$ via the exact solution of \refeq{exactLOSE} in two- and three-body systems extends to large-$\mathrm{A}$ systems, one attempts to design renormalization prescriptions for a \emph{given} many-body approximation, many-body observables thus being renormalization-invariant by construction. While this approach departs from the original scheme pursued in $\spi$EFT, it is meant to serve as a first step towards rooting a successful (yet hypothetic) many-body approximation into an EFT that is suited to a large range of nuclear systems.

\subsection{Set-up of $\mathrm{A}$-body calculations}
\label{subsec:RenoAbody}

\subsubsection{Hamiltonian}
\label{subsubsec:H}

Consider the Hamiltonian
\begin{align}\label{geneH}
  H &\equiv
  \sum_{\vec{p}\sigma}
      \frac{p^{2}}{2m}
      a^\dagger_{\vec{p} \sigma}
      a_{\vec{p} \sigma} \nonumber \\
   &\phantom{=} + \frac{1}{2!} \sum_{\sigma_1 \sigma_2}
   \sum_{\substack{\vec{p}_1 \vec{p}_2 \\ \vec{p}_1^\exprime \vec{p}_2^\exprime}} \
       (2\pi)^3 \delta(\vec{p}_1^\exprime + \vec{p}_2^\exprime - \vec{p}_1 - \vec{p}_2) \ C_0
       \nonumber \\
     &\phantom{= + \frac{1}{2!} \sum \sum \quad}
     a^\dagger_{\vec{p}_1^\exprime \sigma_1}
     a^\dagger_{\vec{p}_2^\exprime \sigma_2}
     a_{\vec{p}_1 \sigma_1}
     a_{\vec{p}_2 \sigma_2}
\end{align}
containing, for convenience, a sole spin-independent two-body interaction parametrised by $C_0$.
All following derivations can be extended, with minimal modifications, to include spin-isospin dependence and three-body interactions (hence including the case of $\spi$EFT at LO). Homogeneous neutron matter would typically be the first system of practical interest, in which case no three-body interaction and no isospin quantum number need to be consider at LO in $\spi$EFT.

\subsubsection{Perturbation theory}
\label{subsubsec:PT}

Approximations presently considered are formulated within MBPT on the basis of the particular partitioning of $H$
\begin{subequations}\label{kineticPartH}
\begin{align}
  H &\equiv H_0 + H_1 \ , \\
  H_0 &\equiv \sum_{\vec{p}\sigma}
      \frac{p^{2}}{2m}
      a^\dagger_{\vec{p} \sigma}
      a_{\vec{p} \sigma} \ , \\
  H_1 &\equiv \frac{1}{2!} \sum_{\sigma_1 \sigma_2}
  \sum_{\substack{\vec{p}_1 \vec{p}_2 \\ \vec{p}_1^\exprime \vec{p}_2^\exprime}} \
  (2\pi)^3 \delta(\vec{p}_1^\exprime + \vec{p}_2^\exprime - \vec{p}_1 - \vec{p}_2) \ C_0
  \nonumber \\
  &\phantom{= + \frac{1}{2!} \sum \sum \quad}
  a^\dagger_{\vec{p}_1^\exprime \sigma_1}
  a^\dagger_{\vec{p}_2^\exprime \sigma_2}
  a_{\vec{p}_1 \sigma_1}
  a_{\vec{p}_2 \sigma_2} \ , \label{IntVertC0}
\end{align}
\end{subequations}
\textit{i.e.}\ the unperturbed Hamiltonian $H_0$ is taken to be the kinetic energy Hamiltonian. Because $H_0$ is invariant under spatial translations, this partitioning is convenient to study homogeneous nuclear matter. In the $\mathrm{A}$-body sector, the unperturbed reference state, \textit{i.e.}\ the ground state of $H_0$, is given by the Slater determinant
\begin{equation}\label{refStateA}
  \ket{\Phi^{\mathrm{A}}_0} \equiv \prod_{i=1}^{\mathrm{A}} a^\dagger_{i} \ket{0} \ ,
\end{equation}
where $\ket{0}$ denotes the particle vacuum. Here, the index $i$ is a shorthand notation to label hole states
$(\sigma_i,\vec{p}_i)$, \textit{i.e.}\ one-body states that are occupied in $\ket{\Phi^{\mathrm{A}}_0}$. Similarly $a$ denotes particle states and Greek letters refer to generic states.

In this framework, observables of interest are obtained from
the $k$-body Green's functions\footnote{For example, the ground-state energy $\mathrm{A}$-body energy is obtained from $G^{(\mathrm{A},1)}$ via the so-called Galitskii-Koltun sum rule. Based on an appropriate choice of the ordering of its time labels, the Lehmann's representation of the $k$-body Green's function
\eqref{kbodyGFdef} further accesses the eigenenergies of all systems whose number of particles is comprised between $\mathrm{A} - k$ and $\mathrm{A} + k$. In practice, one has $k \ll \mathrm{A}$ such that accessible observables are associated to states lying in the neighbourhood of the $\mathrm{A}$-body sector used to set the reference state.} defined as
\begin{multline} \label{kbodyGFdef}
  i^k G^{(\mathrm{A},k)}_{\substack{\mu_1 \dots \mu_k \\ \nu_1 \dots \nu_k}}
  (t_{\mu_1},\dots,t_{\mu_k}, t_{\nu_1}, \dots ,t_{\nu_k}) \equiv  \\
          \frac{
                \bracketop{\Psi^\mathrm{A}_0}{
                              \mathrm{T}\left[
                                  a_{\mu_k}(t_{\mu_k}) \dots a_{\mu_1}(t_{\mu_1})
                                  a^\dagger_{\nu_1}(t_{\nu_1}) \dots a^\dagger_{\nu_k}(t_{\nu_k})
                                          \right]}
                          {\Psi^\mathrm{A}_0}
                }
                {\bracket{\Psi^\mathrm{A}_0}{\Psi^\mathrm{A}_0}} \ ,
\end{multline}
where $\mathrm{T}$ denotes the time-ordering operator, creation and annihilation operators are in the Heisenberg picture, and
$\ket{\Psi^\mathrm{A}_0}$ is the exact ground-state of $H$ connected to $\ket{\Phi^{\mathrm{A}}_0}$
via the adiabatic theorem of Gell-Mann and Low~\cite{Gell-Mann1951}. Exact Green's functions can be
themselves expressed as a sum over the complete set of linked diagrams $\mathcal{G}^{(\mathrm{A},k)}_n$ carrying $k$ incoming
and $k$ outgoing external lines \textit{i.e.}\
\begin{multline} \label{kGFt_sumDiags}
  i^k G^{(\mathrm{A},k)}_{\substack{\mu_1 \dots \mu_k \\ \nu_1 \dots \nu_k}}
  (t_{\mu_1} \dots t_{\mu_k} ,t_{\nu_1} \dots t_{\nu_k})
    =  \\
      \sum^{+\infty}_{n=0} \sum_{\mathcal{G}^{(\mathrm{A},k)}_n \in \mathcal{S}^{(\mathrm{A},k)}_{\text{Exact}}} \mathcal{A}^{\mathcal{G}^{(\mathrm{A},k)}_n}_{\substack{\mu_1 \dots \mu_k \\ \nu_1 \dots \nu_k}}(t_{\mu_1} \dots t_{\mu_k} ,t_{\nu_1} \dots t_{\nu_k})
\end{multline}
where $\mathcal{S}^{(\mathrm{A},k)}_{\text{Exact}}$ denotes the complete set of diagrams contributing to the $k$-body Green's function, $n$ the number of interaction vertices and $\mathcal{A}^{\mathcal{G}^{(\mathrm{A},k)}_n}_{\substack{\mu_1 \dots \mu_k \\ \nu_1 \dots \nu_k}}
(t_{\mu_1} \dots t_{\mu_k} ,t_{\nu_1} \dots t_{\nu_k})$ the amplitude associated to
$\mathcal{G}^{(\mathrm{A},k)}_n$.

Below, Green's functions are expressed in the energy
representation for convenience. A similar expression to \refeq{kGFt_sumDiags} holds
in that case. Further, the amplitudes can be expressed in terms of the particle and hole parts of the unperturbed
one-body Green's function associated to the partitioning \eqref{kineticPartH} and the reference state \eqref{refStateA}, respectively reading as
\begin{subequations}
\begin{align}
  iG^{(\mathrm{A},1)0+}_{\mu\nu} &\equiv
        i \sum_{a>\mathrm{A}} \frac{\delta_{\mu a} \delta_{\mu\nu}}{\omega - \frac{\vec{p}_a^2}{2m} + i\eta} \ , \label{propP}\\
  iG^{(\mathrm{A},1)0-}_{\mu\nu} &\equiv
        i \sum_{i=1}^{\mathrm{A}} \frac{\delta_{\mu i} \delta_{\mu\nu}}{\omega - \frac{\vec{p}_i^2}{2m} - i\eta} \ , \label{propH}
\end{align}
\end{subequations}
where the limit $\eta \to 0^+$ is implicit. Explicitly employing the two parts of the one-body Green's function relates to using  time-ordered diagrams. Focusing on one diagram $\mathcal{G}^{(\mathrm{A},k)}_n$
belonging to $\mathcal{S}^{(\mathrm{A},k)}_{\text{Exact}}$,
the associated amplitude reads as
\begin{align}\label{AmplitudePH}
  \mathcal{A}^{\mathcal{G}^{(\mathrm{A},k)}_n}_{\substack{\mu_1 \dots \mu_k \\ \nu_1 \dots \nu_k}}
  (\omega_{\mu_1} \dots \omega_{\mu_k} ,\omega_{\nu_1} \dots \omega_{\nu_k})
    &= \phantom{(-1)^\sigma} \nonumber \\
      (-1)^\sigma \frac{(-i)^n}{n!} \sum_{\lambda}
                    \frac{h^{22}_{\lambda \dots \lambda}}{(2!)^2} \dots
                    \frac{h^{22}_{\lambda \dots \lambda}}{(2!)^2} \nonumber \\
                    \int \frac{\mathrm{d}\omega_\lambda}{2\pi} \dots
                    \prod_{i=1}^{n}
                    2\pi \delta\left( \pm \omega_\lambda \dots - \omega_\mu \dots + \omega_\nu \dots \right)&
                    \nonumber \\
                      \times \prod_{e\in I^{+}} iG^{(\mathrm{A},1)0+}_{\lambda\lambda}(\omega_\lambda)
                      \prod_{e\in I^{-}} iG^{(\mathrm{A},1)0-}_{\lambda\lambda}(\omega_\lambda)& \nonumber \\
                      \times \prod_{e\in E_{in}^{+}} iG^{(\mathrm{A},1)0+}_{\lambda\nu}(\omega_{\nu})
                      \prod_{e\in E_{in}^{-}} iG^{(\mathrm{A},1)0-}_{\lambda\nu}(\omega_{\nu})& \nonumber \\
                      \times \prod_{e\in E_{out}^{+}} iG^{(\mathrm{A},1)0+}_{\mu\lambda}(\omega_{\mu})
                      \prod_{e\in E_{out}^{-}} iG^{(\mathrm{A},1)0-}_{\mu\lambda}(\omega_{\mu})&
 \ ,
\end{align}
where $\sigma$ is an integer, $\lambda$, $\mu$ and $\nu$ denote generic single-particle labels
for internal, external outgoing and external incoming lines, respectively. Additionally,
$I^{+}$ ($I^{-}$) denotes the set of internal particle (hole) lines,
$E_{in}^{+}$ ($E_{in}^{-}$) the set of external incoming particle (hole) lines and
$E_{out}^{+}$ ($E_{out}^{-}$) the set of external outgoing particle (hole) lines.

Many-body approximations considered here consist of selecting a subset
$\mathcal{S}^{(\mathrm{A},k)}_{\text{MB}} \subset \mathcal{S}^{(\mathrm{A},k)}_{\text{Exact}}$
so that the approximated $k$-body Green's function reads as
\begin{multline}\label{Approx_kGF_PH}
  i^k G^{(\mathrm{A},k)\text{MB}}_{\substack{\mu_1 \dots \mu_k \\ \nu_1 \dots \nu_k}}
  (\omega_{\mu_1} \dots \omega_{\mu_k} ,\omega_{\nu_1} \dots \omega_{\nu_k})
    \equiv \\
      \sum_{\mathcal{G}^{(\mathrm{A},k)} \in \mathcal{S}^{(\mathrm{A},k)}_{\text{MB}} }
       \mathcal{A}^{\mathcal{G}^{(\mathrm{A},k)}}_{\substack{\mu_1 \dots \mu_k \\ \nu_1 \dots \nu_k}}
       (\omega_{\mu_1} \dots \omega_{\mu_k} ,\omega_{\nu_1} \dots \omega_{\nu_k})
 \ .
\end{multline}

\subsubsection{Regularization}
\label{subsubsec:reg}

When considering local interactions, as those derived in any EFT, the
amplitude $\mathcal{A}^{\mathcal{G}^{(\mathrm{A},k)}}_{\substack{\mu_1 \dots \mu_k \\ \nu_1 \dots \nu_k}}$
 contains, in general, ultraviolet (UV) divergences requiring the introduction of
a regularisation to suppress the high-momentum modes. In this work, the regularisation
is introduced via a momentum regulator $v_\Lambda(q)$ satisfying
\begin{subequations}\label{regMomSep}
\begin{align}
  \lim_{\Lambda \to +\infty} v_\Lambda(q) &= 1 \ , \label{limitReg} \\
  \lim_{q \to +\infty} v_\Lambda(q) &= 0 \ , \label{limitUVsupp} \\
  v_\Lambda(0) &= 1 \ , \label{normalisation} \\
  \int^{+\infty}_{0} \mathrm{d}q \ v^2_\Lambda(q) &< +\infty  \, , \label{UVwelldefined}
\end{align}
\end{subequations}
where $\Lambda$ represents the characteristic cut-off scale beyond which high-momentum modes
are suppressed and $\vec{q}$ denotes the relative momentum between two nucleons.
Eventually, one is interested in the limit where $\Lambda \gg Q$
with $Q$ the scale characterising the low-energy observables of interest.
The limit of large $\Lambda$ ensures that the evaluation of observables is independent of the high-energy physics.
The difference between the actual high-momentum interaction (which is unknown)
and the regularised field theory one (which is arbitrary) is then captured effectively in its
parameters referred to as low energy constants (LECs). Correspondingly, the LECs explicitly depend on the regulator
$v_\Lambda$ and in particular on the cut-off $\Lambda$. Eventually, the Hamiltonian to be actually considered reads as
\begin{align}\label{RegH}
  H &\equiv
  \sum_{\vec{p}\sigma}
      \frac{p^{2}}{2m}
      a^\dagger_{\vec{p} \sigma}
      a_{\vec{p} \sigma} \nonumber \\
   &\phantom{=} + \frac{1}{2!} \sum_{\sigma_1 \sigma_2}
   \sum_{\substack{\vec{p}_1 \vec{p}_2 \\ \vec{p}_1^\exprime \vec{p}_2^\exprime}} \
       (2\pi)^3 \delta(\vec{p}_1^\exprime + \vec{p}_2^\exprime - \vec{p}_1 - \vec{p}_2) \ C_0(\Lambda) \
       \nonumber \\
     &\phantom{= + \frac{1}{2!} \sum \sum \quad}
     \times v_\Lambda(\vec{q}) v_\Lambda(\vec{q}^\exprime) \
     a^\dagger_{\vec{p}_1^\exprime \sigma_1}
     a^\dagger_{\vec{p}_2^\exprime \sigma_2}
     a_{\vec{p}_1 \sigma_1}
     a_{\vec{p}_2 \sigma_2} \ ,
\end{align}
where the dependence of the LEC $C_0(\Lambda)$ on the regularisation is made explicit and where incoming and outgoing relative momenta are respectively defined as
\begin{subequations}
\begin{align}
  \vec{q} &\equiv \frac{\vec{p}_1 - \vec{p}_2}{2} \ , \\
  \vec{q}^\exprime &\equiv \frac{\vec{p}_1^\exprime - \vec{p}_2^\exprime}{2} \ .
\end{align}
\end{subequations}
To make the cancellation of UV divergences explicit, LECs are usually decomposed, for a given
renormalization scheme, into a renormalized component and a counterterm \cite{Collins1984},
\textit{i.e.}\ in the case of \eqref{RegH} as
\begin{equation}
  C_0(\Lambda) = C_0^R + \delta C_0(\Lambda) \ ,
\end{equation}
where $C^R_0$ is the finite renormalized LEC and $\delta C_0(\Lambda)$ encapsulates all counterterms
that will be necessary to cancel UV divergences.
For a given set of diagrams $\mathcal{S}^{(\mathrm{A},k)}_{\text{MB}}$ with renormalized vertices $C^R_0$,
the aim is thus to derive an additional set of diagrams
$\mathcal{S}^{(\mathrm{A},k)}_{\text{MB,ct}}$ with renormalized and counterterm vertices
(to be derived as well) such that the total sum
\begin{multline}\label{Approx_kGF_PH_CT}
  i^k G^{(\mathrm{A},k)\text{RMB}}_{\substack{\mu_1 \dots \mu_k \\ \nu_1 \dots \nu_k}}
  (\omega_{\mu_1} \dots \omega_{\mu_k} ,\omega_{\nu_1} \dots \omega_{\nu_k})
    \equiv \\
      \sum_{\mathcal{G}^{(\mathrm{A},k)} \in \mathcal{S}^{(\mathrm{A},k)}_{\text{MB}}
      \cup \mathcal{S}^{(\mathrm{A},k)}_{\text{MB,ct}} }
       \mathcal{A}^{\mathcal{G}^{(\mathrm{A},k)}}_{\substack{\mu_1 \dots \mu_k \\ \nu_1 \dots \nu_k}}
       (\omega_{\mu_1} \dots \omega_{\mu_k} ,\omega_{\nu_1} \dots \omega_{\nu_k})
\end{multline}
converges to a finite value when $\Lambda \gg Q$.

In the perturbation theory employing the particle vacuum $\ket{0}$ as a reference state,
the counterterms needed to achieve renormalization can be identified systematically via the application of the BPHZ
theorem~\cite{Bogoliubov1957,Hepp1966,Hahn1968,Zimmermann1968,Zimmermann1969,Lowenstein1975}.
Applying a similar procedure to time-ordered diagrams stemming from a perturbative expansion around an in-medium reference state $\ket{\Phi^{\mathrm{A}}_0}$ poses some challenges. One must overcome these challenges given that perturbation theory, or any other expansion many-body method for that matter, can only be efficiently formulated around an in-medium reference state as soon as one targets systems containing more than a few particles. One example of apparent difficulty relates to the fact that the renormalization procedure, if possible, might have to be achieved for each $\mathrm{A}$ given that the particle and hole propagators do depend on $\mathrm{A}$. In numerical calculations, obervables used to renormalize the LECs would have to be computed for each $\mathrm{A}$. This situation would penalise both numerical efficiency and the predictive power of the theory.

Therefore, the goal of the present manuscript is to design a renormalization procedure for observables computed on the basis of a given MBPT approximation such that necessary $k$-body counterterms, hopefully limited to small $k$, are independent of $\mathrm{A} \geq k$.

\section{Basic tools}
\label{sec:BasicTheo}

Weinberg's asymptotic theorem~\cite{Weinberg1960} and the BPHZ~\cite{Bogoliubov1957,Hepp1966,Hahn1968,Zimmermann1968,Zimmermann1969,Lowenstein1975} theorem represent key tools to develop the renormalization procedure exposed in Sec.~\ref{sec:RenoMB} below. Consequently, the two theorems are briefly recalled in the present section.

\subsection{Weinberg's Asymptotic theorem}

The UV convergence of the amplitude of a time-ordered diagram is analysed as the convergence problem of a multidimensional
integral of a multivariate function. The main ingredients to understand Weinberg's asymptotic theorem
are presently introduced for a generic multivariate function. Furthermore, the theorem is recast
in terms of diagrams. For the complete proof and further discussion on Weinberg's asymptotic theorem, see Ref.~\cite{Weinberg1960}.

\subsubsection{Asymptotic coefficients and multidimensional integrals}

Let us first introduce the definition of asymptotic coefficients (provided they exist) of a function
$f : \mathbb{R}^n \to \mathbb{C}$ as given in~\cite{Weinberg1960}.
For any vector subspace\footnote{For convenience,
in this section, the bracket notation $\{ \dots \}$ denotes the vector space spanned by the set of vectors
considered.} $S = \left\{\vec{L}_1, \dots, \vec{L}_m \right\}$ of $\mathbb{R}^n$
with $m \leq n$ and $\vec{L}_1, \dots, \vec{L}_m$ being $m$ independent $\mathbb{R}^n$-vectors,
and any compact region $W \subset \mathbb{R}^n$,
the asymptotic coefficients are defined as the numbers
$\alpha\left( \left\{ \vec{L}_1, \dots, \vec{L}_r \right\}\right)$
and $\beta\left( \left\{ \vec{L}_1, \dots, \vec{L}_r \right\}\right)$ (with $1 \leq r \leq m$) such that
for any $\vec{C} \in W$
\begin{multline}\label{Def_AsymptoticCoeffs}
  f\left( \eta_1 \dots \eta_m \vec{L}_1 + \eta_2 \dots \eta_m \vec{L}_2 + \dots + \eta_m \vec{L}_m + \vec{C} \right)
  = \\
    O\left( \eta_1^{\alpha\left(\left\{\vec{L}_1\right\}\right)}
            \left( \ln \eta_1 \right)^{\beta\left(\left\{\vec{L}_1\right\}\right)} \right.\\
            \left. \eta_2^{\alpha\left(\left\{\vec{L}_1,\vec{L}_2\right\}\right)}
            \left( \ln \eta_2 \right)^{\beta\left(\left\{\vec{L}_1,\vec{L}_2\right\}\right)}
            \times  \dots \right.\\
            \left. \dots \times
            \eta_m^{\alpha\left(\left\{\vec{L}_1, \dots, \vec{L}_m\right\}\right)}
            \left( \ln \eta_m \right)^{\beta\left(\left\{\vec{L}_1, \dots, \vec{L}_m\right\}\right)}
      \right) \ ,
\end{multline}
if $\eta_1 \dots \eta_m$ go independently to infinity.
The asymptotic coefficients $\alpha\left( S \right)$
and $\beta\left( S \right)$ can be interpreted as the asymptotic
coefficients $\alpha\left( \left\{ \vec{L} \right\}\right)$ and $\beta\left(\left\{ \vec{L} \right\}\right)$
for $\vec{L}$ a 'typical' vector in $S$ \textit{i.e.}\ fixing $\eta_1 \dots \eta_{m-1}$ sufficiently large and $\vec{C} \in W$,
\begin{multline}\label{Typical_AsymptoticCoeff_S}
  f\left( \left[\eta_1 \dots \eta_{m-1} \vec{L}_1
          + \eta_2 \dots \eta_{m-1} \vec{L}_2 + \dots \right.\right. \\
          \left.\left. \dots + \eta_{m-1} \vec{L}_{m-1}
          + \vec{L}_m
          \right]  \eta_m
          + \vec{C}
          \right)
          = \\
          O\left( \eta_m^{\alpha\left(S\right)} \left( \ln \eta_m \right)^{\beta\left(S\right)} \right) \ ,
\end{multline}
when $\eta_m$ goes to infinity.

Considering now integrals of a function $f$, the integration
along the directions $\vec{L}_1, \dots, \vec{L}_r$ is defined as
\begin{multline}
  f_{\vec{L}_1, \dots, \vec{L}_r}\left( \vec{X}\right)
    \equiv \\
    \int_{-\infty}^{+\infty} \mathrm{d}y_1 \ \dots
            \int_{-\infty}^{+\infty} \mathrm{d}y_r \
                f\left( \vec{X} + y_1 \vec{L}_1 + \dots + y_r \vec{L}_r \right) \ ,
\end{multline}
where $\vec{X}$ is a vector in $\mathbb{R}^n$.
Thanks to Fubini's theorem, if $f_{\vec{L}_1, \dots, \vec{L}_r}\left( \vec{X}\right)$ exists (in the sense
that the integral is absolutely convergent), the integration depends only on the vector space
$I = \left\{ \vec{L}_1, \dots, \vec{L}_r \right\}$ so that one defines
\begin{equation}
  f_{I}\left(\vec{X}\right) \equiv \int_{\vec{Y} \in I} \mathrm{d}^r\vec{Y} \ f(\vec{X} + \vec{Y})
  \equiv f_{\vec{L}_1, \dots, \vec{L}_r}\left( \vec{X} \right) \ .
\end{equation}
Furthermore, $f_{I}\left(\vec{X}\right)$ depends only on the projection of $\vec{X}$ along $I$\footnote{Any
component of $\vec{X}$ in $I$ can be absorbed in $\vec{Y}$ by a change of variable in the integral on $I$.}.
Choosing a subspace $E$ such that $\mathbb{R}^n = I \oplus E$, the domain of definition of the function
$f_{I}\left(\vec{X}\right)$ can be restricted to $E$.

In the case of the amplitude of a time-ordered diagram, the general integrand depends on $(\omega_1, \vec{p}_1, \dots)$ and
is integrated on the internal energies and momenta.
Therefore in this case, $I$ denotes the vector space of internal (one-body) energies and momenta whereas
the vector space $E$ denotes the space of external (one-body) energies and momenta\footnote{Finite sum on spin-isospin
indices are omitted here as they introduce only a finite linear combination of integrals on energies and momenta
so that the conclusion on the UV behaviour is not impacted.}. The general vector space $\mathbb{R}^n = I \oplus E$
denotes the vector space of all (one-body) energies and momenta (internal and external).

As an example, the asymptotic coefficients $\alpha\left(S\right)$ of the in-vacuum propagator
\begin{equation}\label{unperturbedVacProp}
  iG^{(0,1)0}_{\vec{p}\sigma}(\omega) = \frac{i}{\omega - \frac{p^2}{2m} + i \eta}
\end{equation}
are now extracted. The in-vacuum propagator $iG^{(0,1)0}_{\vec{p}\sigma}(\omega)$ is interpreted
as a multivariate function on the vector space $\mathbb{R}^4 = \{ \vec{e}_\omega, \vec{e}_{p_x}, \vec{e}_{p_y}, \vec{e}_{p_z}\}$,
so that\footnote{The label $\sigma$ is just considered as a fixed parameter so that it is dropped
in the definition of $f(\omega \vec{e}_\omega + p_x \vec{e}_{p_x} + p_y \vec{e}_{p_y} + p_z \vec{e}_{p_z})$.}
\begin{equation}
  f(\omega \vec{e}_\omega + p_x \vec{e}_{p_x} + p_y \vec{e}_{p_y} + p_z \vec{e}_{p_z} )
    \equiv iG^{(0,1)0}_{\vec{p}\sigma}(\omega) \ .
\end{equation}
One can show that, in this case, the asymptotic coefficients of the in-vacuum propagator $\alpha^0$
read as
\begin{equation}
 \alpha^0\left(S\right) =
  \begin{cases}
    -1 &\text{ if } S = \{ \vec{e}_\omega\} \\
    -2 &\text{ if } S = \{ \vec{L} \} \text{ with } \vec{L} \notin \{\vec{e}_\omega\} \\
    -2 &\text{ if } \dim{S} \geq 2
  \end{cases}
  \ .
\end{equation}

\subsubsection{Asymptotic theorem}
\label{subsubsec:AsTheo}

With all the notations introduced before, the general asymptotic theorem follows

\emph{If a function $f\left(\vec{X}\right)$ possesses asymptotic coefficients $\alpha\left(S\right),\beta\left(S\right)$
for any non-null subspace $S \subset \mathbb{R}^n$, if $f(\vec{X})$ is integrable for any finite region of $\mathbb{R}^n$ and if $D_I < 0$ where
\begin{equation}
  D_I \equiv \max_{S^\exprime \subseteq I} \left[ \alpha\left(S^\exprime\right) + \dim{S^\exprime} \right] \ ,
\end{equation}
then $f_I\left(\vec{X}\right)$ exists i.e.\ is absolutely convergent.}

To apply this theorem to a time-ordered diagram
$\mathcal{G}$, it is reformulated in terms of sub-diagrams as detailed in Ref.~\cite{Weinberg1960}.
This is done by associating to any sub-space of integration $S^\exprime \subseteq I$
a sub-diagram $\gamma \subseteq \mathcal{G}$ made of internal lines.
In particular $S^\exprime = I$ corresponds to the sub-diagram $\gamma$ made of $\mathcal{G}$ itself
without its external lines.
The quantity $\alpha\left(S^\exprime\right) + \dim{S^\exprime}$ corresponds to
the superficial degree of divergence of the associated sub-diagram $\gamma$ of $\mathcal{G}$
and is denoted $D(\gamma)$. For a diagram $\mathcal{G}^{(0,k)}_n$ where lines denote the unperturbed in-vacuum
propagator \eqref{unperturbedVacProp}, $n$ denotes the number of vertices
\eqref{IntVertC0} and $k$ is the number of incoming (and of outgoing) external lines,
\begin{equation}\label{Simplified_SuperficialD}
  D\left(\mathcal{G}^{(0,k)}_n\right) = 5 - 3k + n \ .
\end{equation}
Having $D_I < 0$ is thus equivalent to having $D(\gamma) < 0$ for all $\gamma \subseteq \mathcal{G}$
made of internal lines.
Consequently, from Weinberg's asymptotic theorem, the well-known power-counting theorem follows.
\emph{The amplitude associated to $\mathcal{G}$ is finite if $D(\gamma) < 0$
for any sub-diagram $\gamma \subseteq \mathcal{G}$ made of internal lines of $\mathcal{G}$.}

Weinberg's asymptotic theorem is very powerful as it proves the convergence of the
amplitude associated to a diagram
\emph{with the sole knowledge of the asymptotic coefficients $\alpha\left(S\right)$ associated to the propagator}.
This particular property is fundamental for the development of the
renormalization prescription exposed in \refsec{sec:RenoMB}.

\subsection{Subtractions and BPHZ theorem}
\label{subsec:BPHZ}

Whilst the asymptotic theorem allows to prove the convergence of the calculations
with adequate counterterms, BPHZ theorem offers a systematic way to generate
sufficient counterterms to lower the superficial degree of divergence of the internal
sub-diagrams $\gamma \subseteq \mathcal{G}$. The procedure to do so is now briefly detailed.

\subsubsection{Definitions}

A sub-diagram $\gamma$ of $\mathcal{G}$ is defined as a subset of lines and vertices contained
in $\mathcal{G}$ where end points of the lines of $\gamma$ belong to its vertices.
A diagram $\mathcal{G}_1$ is said to be included in $\mathcal{G}_2$ and denoted as $\mathcal{G}_1 \subseteq \mathcal{G}_2$
if their set of lines verify the same inclusion relation.
In particular, a sub-diagram $\gamma$ of $\mathcal{G}$ verifies $\gamma \subseteq \mathcal{G}$.
The sub-diagram generated by the intersection of lines of two sub-diagrams $\gamma_1$ and $\gamma_2$
defines a sub-diagram $\gamma$ and is denoted as
\begin{equation}
 \gamma \equiv \gamma_1 \cap \gamma_2 \ .
\end{equation}
Two sub-diagrams  $\gamma_1$ and $\gamma_2$ that have neither lines nor vertices in common
are said to be \emph{disjoint} and the result is denoted as
\begin{equation}
 \gamma_1 \cap \gamma_2 = \emptyset \ .
\end{equation}
If neither $\gamma_1 \subseteq \gamma_2$ nor $\gamma_2 \subseteq \gamma_1$ and $\gamma_1 \cap \gamma_2 \neq \emptyset$
they are said to be \emph{overlapping}. Otherwise they are said to be non-overlapping.
For a set of non-overlapping sub-diagrams $\gamma_1,\gamma_2,\ldots,\gamma_n$ of $\mathcal{G}$,
the \emph{reduced diagram} $\mathcal{G} \setminus \{\gamma_1,\gamma_2,\ldots,\gamma_n\}$
is defined by the diagram resulting from $\mathcal{G}$ after contracting all lines of
$\gamma_1,\gamma_2,\ldots,\gamma_n$ to a point.

The amplitude associated to $\mathcal{G}$ is denoted as $\mathcal{A}^{\mathcal{G}}$.
Given a set of mutually disjoint sub-diagrams $\gamma_1,\gamma_2,\ldots,\gamma_n$ of $\mathcal{G}$, the
corresponding amplitude is expressed as the product of the amplitudes $\mathcal{A}^{\gamma_j}$ of the sub-diagrams
 while the remainder is denoted as
$\mathcal{A}^{\mathcal{G} \setminus \{\gamma_1,\ldots,\gamma_n \} }$ such that
\begin{equation}
  \mathcal{A}^{\mathcal{G}} = \mathcal{A}^{\mathcal{G} \setminus \{\gamma_1,\ldots,\gamma_n \} }
  \prod_{j=1}^{n} \mathcal{A}^{\gamma_j}. \label{factorization}
\end{equation}

A sub-diagram $\gamma$ of $\mathcal{G}$ is referred to as a \emph{renormalization part}
if it is a one-particle irreducible (1PI) diagram with a superficial degree
of divergence greater than or equal to $0$, \textit{i.e.}\ if
\begin{equation}\label{Def_renormalization_part}
 D(\gamma) \geq 0 \ .
\end{equation}

\subsubsection{Recursive subtractions of UV divergences}

The BPHZ procedure defines recursively a renormalized
amplitude $R^{\mathcal{G}}$ associated to the diagram $\mathcal{G}$~\cite{Bogoliubov1957,Hepp1966}.
If the amplitude associated to $\mathcal{G}$ is convergent to begin with, then
\begin{equation}
  R^{\mathcal{G}} \equiv \mathcal{A}^{\mathcal{G}} \ .
\end{equation}
If the diagram does not contain any renormalization part but is superficially
divergent, it is called \emph{primitively divergent}. In that case the
renormalized amplitude is defined by
\begin{equation}\label{primDiv_renoAmplitude}
  R^{\mathcal{G}} \equiv (1-t_{\mathcal{G}}) \mathcal{A}^{\mathcal{G}} \ ,
\end{equation}
where $t_{\mathcal{G}}$ is the operator of the Taylor
expansion with respect to the external momenta\footnote{The renormalization point is chosen here
to be $0$ but could be chosen arbitrarily.} around $0$ up to the
order of the superficial degree of divergence $D(\gamma)$ of the diagram, \textit{i.e.}\
\begin{multline}\label{Def_taylorSub}
t_{\gamma} \mathcal{A}^{\gamma}_{\substack{p_{1}, \dots, p_{k} \\ p_{1}^\exprime, \dots, p_{k}^\exprime}}
  \equiv
          \sum_{j=0}^{D(\gamma)}
          \frac{1}{j!}
          \sum_{\substack{s_{1} + \dots + s_{k}^\exprime \ge 0 \\
                                       s_{1} +\dots+ s_{k}^\exprime =j}} \\
                  \left .
                  \frac{\partial^{j}\mathcal{A}^{\gamma}}
                       {\partial p_{1}^{s_{1}} \cdots   \partial p_{k}^{s_{k}}
                        \partial p_{1}^{\exprime s_{1}^\exprime} \cdots   \partial p_{k}^{\exprime s_{k}^\exprime}
                       }
                  \right |_{ p_{1} = \dots = p_{k}^\exprime = 0 }
                  p_{1}^{s_{1}}  \cdots p_{k}^{\exprime s_{k}^\exprime} \vspace{2mm} \ .
\end{multline}
If $\mathcal{G}$ is superficially divergent and contains
divergent sub-diagrams, the renormalized amplitude is defined recursively as
\begin{equation}\label{recursionRelation}
  R^{\mathcal{G}} \equiv
                      (1-t_{\mathcal{G}})\bar{R}^{\mathcal{G}} \ ,
\end{equation}
where $\bar{R}^{\mathcal{G}}$ corresponds to the amplitude where
all sub-divergences have already been subtracted.
The subtraction by the Taylor operator $t_{\gamma}$ corresponds to adding the amplitude
associated to a diagram where the divergent sub-diagram $\gamma$ has been replaced
by a so-called \emph{counterterm} vertex.
Here $t_{\gamma}$ corresponds to
a zero momentum subtraction of UV divergences. Different subtractions can be used
by modifying the definition of $t_{\gamma}$~\cite{Collins1984}.

\subsubsection{Forest formula}

\begin{figure}
\centering
\parbox{50pt}{\begin{fmffile}{Figures/diags/Ex_iForest_3loop}
   \begin{fmfgraph*}(80,120)
     \fmfkeep{Ex_iForest_3loop}
   \fmfbottom{b1,b2,b3,b4} \fmftop{t1,t2,t3,t4}
   \fmf{fermion, tension=1.5}{b2,v1}
   \fmf{fermion, tension=1.5}{b3,v1}
   \fmf{fermion, left=0.75}{v1,v2}
   \fmf{fermion, right=0.75}{v1,v2}
   \fmf{fermion, left=0.75}{v2,v3}
   \fmf{fermion, right=0.75}{v2,v3}
   \fmf{fermion, left=0.75}{v3,v4}
   \fmf{fermion, right=0.75}{v3,v4}
   \fmf{fermion, tension=1.5}{v4,t2}
   \fmf{fermion, tension=1.5}{v4,t3}
   \fmfv{d.shape=circle,d.filled=full,d.size=3thick}{v1,v2,v3,v4}

   \fmffreeze

   \fmfcmd{%
      save loc, csmin, csmax, clmin, clmax;
      forsuffixes $ = 1, 2, 3:
        (loc.$.x, loc.$.y) = vloc(__v.$);
      endfor
      csmax.x = max (loc1x, loc2x) + .2w;
      csmax.y = max (loc1y, loc2y) + .05h;
      csmin.x = min (loc1x, loc2x) - .2w;
      csmin.y = min (loc1y, loc2y) - .05h;
      clmax.x = max (loc1x, loc3x) + .4w;
      clmax.y = max (loc1y, loc3y) + .1h;
      clmin.x = min (loc1x, loc3x) - .4w;
      clmin.y = min (loc1y, loc3y) - .1h;
    }
    \fmfi{dashes,width=thin}{%
      (csmin.x,csmin.y) -- (csmax.x,csmin.y)
      -- (csmax.x,csmax.y) -- (csmin.x,csmax.y) -- cycle
    }
    \fmfi{dashes,width=thin}{%
      (clmin.x,clmin.y) -- (clmax.x,clmin.y)
      -- (clmax.x,clmax.y) -- (clmin.x,clmax.y) -- cycle
    }
    \fmfiv{label=$\gamma_1$}%
          {(0.95csmax.x,csmax.y)}
    \fmfiv{label=$\gamma_2$}%
          {(0.95clmax.x,clmax.y)}
  \end{fmfgraph*}
\end{fmffile}}
\hspace{2cm}
\parbox{25pt}{\begin{fmffile}{Figures/diags/Ex_ct_3loop}
   \begin{fmfgraph*}(40,120)
     \fmfkeep{Ex_ct_3loop}
   \fmfbottom{b1,b2,b3,b4} \fmftop{t1,t2,t3,t4}
   \fmf{fermion, tension=2}{b2,v1}
   \fmf{fermion, tension=2}{b3,v1}

   \fmf{fermion, left=0.5}{v1,v2}
   \fmf{fermion, right=0.5}{v1,v2}

   \fmf{fermion, tension=2}{v2,t2}
   \fmf{fermion, tension=2}{v2,t3}

   \fmfv{d.shape=circle,d.filled=empty,d.size=4thick}{v1}
   \fmfv{d.shape=circle,d.filled=full,d.size=3thick}{v2}
  \end{fmfgraph*}
\end{fmffile}}
\hspace{2cm}
\parbox{25pt}{\begin{fmffile}{Figures/diags/Ex_Hasse_3loop}
   \begin{fmfgraph*}(40,120)
     \fmfkeep{Ex_Hasse_3loop}
   \fmfbottom{b1} \fmftop{t1}
   \fmf{phantom}{b1,v1}
   \fmf{fermion, tag=1}{v1,v2}
   \fmf{phantom}{v2,t1}

   \fmfv{d.shape=square,d.filled=full,d.size=3thick,
          label=$\gamma_1$,l.angle=180,l.dist=7thick}{v1}
   \fmfv{d.shape=square,d.filled=full,d.size=3thick,
          label=$\gamma_2$,l.angle=180,l.dist=7thick}{v2}

    \fmfposition
    \fmfipath{p[]}
    \fmfiset{p1}{vpath1(__v1,__v2)}
    \fmfiv{label=$\subseteq$,l.dist=4thick,l.angle=0}{point length(p1)/2 of p1}
  \end{fmfgraph*}
\end{fmffile}}
\caption{On the left panel, an example of an i-forest (depicted by boxes) for a three-loop ladder diagram
contributing to $G^{(0,2)}$.
The middle panel pictures the diagram with the counterterm associated to this i-forest (the vertex with an empty
circle denotes the two-body counterterm appearing for this particular i-forest). The associated Hasse diagram is depicted on the right panel.}
\label{Example_BPHZ_subtraction}
\end{figure}
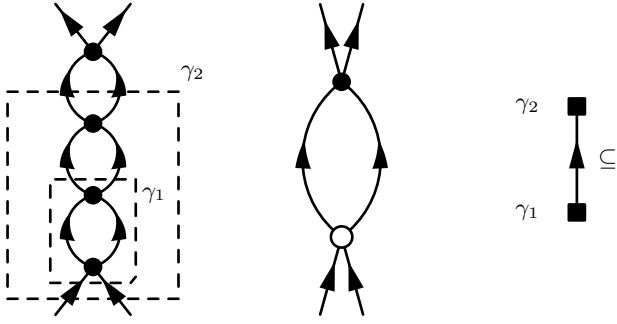

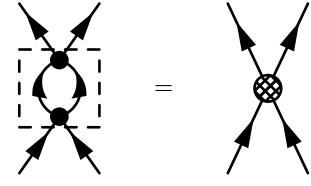
\begin{figure}
\centering
\hspace{0.5cm}
\parbox{50pt}{
\begin{fmffile}{Figures/diags/UV_LoopNoMedium_BPHZ_boxed}
\begin{fmfgraph*}(30,80)
  \fmfbottom{i1,i2}
  \fmftop{o1,o2}
  \fmf{fermion}{i1,v1}
  \fmf{fermion}{i2,v1}
  \fmf{fermion}{v2,o1}
  \fmf{fermion}{v2,o2}
  \fmf{fermion, left=0.75}{v1,v2}
  \fmf{fermion, right=0.75}{v1,v2}
  \fmfv{d.shape=circle,d.filled=full,d.size=3thick}{v1}
  \fmfv{d.shape=circle,d.filled=full,d.size=3thick}{v2}
\fmffreeze
\fmfcmd{%
   save loc, csmin, csmax;
   forsuffixes $ = 1, 2:
     (loc.$.x, loc.$.y) = vloc(__v.$);
   endfor
   csmax.x = max (loc1x, loc2x) + .5w;
   csmax.y = max (loc1y, loc2y) + .05h;
   csmin.x = min (loc1x, loc2x) - .5w;
   csmin.y = min (loc1y, loc2y) - .05h;
 }
 \fmfi{dashes,width=thin}{%
   (csmin.x,csmin.y) -- (csmax.x,csmin.y)
   -- (csmax.x,csmax.y) -- (csmin.x,csmax.y) -- cycle
 }
\end{fmfgraph*}
\end{fmffile}
}
\hspace{-0.2cm}
=
\hspace{0.5cm}
\parbox{50pt}{
\begin{fmffile}{Figures/diags/UV_LoopDiagct_BPHZ_explicit}
\begin{fmfgraph*}(30,80)
\fmfbottom{i1,i2}
\fmftop{o1,o2}
\fmf{fermion}{i1,v1}
\fmf{fermion}{i2,v1}
\fmf{fermion}{v1,o1}
\fmf{fermion}{v1,o2}
\fmfv{d.shape=circle,d.filled=hatched,d.size=5thick}{v1}
\end{fmfgraph*}
\end{fmffile}
}
\caption{Representation, in the case of a one-loop diagram, of the amplitude $\Omega^{\mathscr{F}}$
both with the i-forest pictured on the original diagram and with an explicit counter-term vertex.
The filled dot represents $C^R_0$ whereas the hatched vertex represents the counterterm associated to
the i-forest represented by a box on the left-hand side diagram.}
\label{1loop_iForestNotation}
\end{figure}

The recursion relation~\eqref{recursionRelation} was solved explicitly by the \emph{forest formula}~\cite{Zimmermann1969} that is based on the concept of \emph{i-forest} (for inclusion-forest). An i-forest is defined as any set of sub-diagrams (including the empty set) that are mutually non-overlapping. This way, the Hasse diagram\footnote{A Hasse diagram associated to an order relation is a diagrammatic representation of the ordering between the objects considered.}, for the order relation $\subseteq$ on the mutually non-overlapping sub-diagrams, represents a forest \textit{i.e.}\ a set of disconnected trees (see the right panel of \reffig{Example_BPHZ_subtraction} for an example where the Hasse diagram of the i-forest is made of only one tree.) An i-forest is said to be connected if its Hasse diagram is connected. A connected i-forest is also referred to as an \emph{i-tree}. As for usual forests, an i-forest can be decomposed as the set of its connected components (\textit{i.e.}\ as a set of i-trees). An i-forest is usually depicted by drawing boxes around the sub-diagrams as illustrated in the left panel of \reffig{Example_BPHZ_subtraction}. The boxes are, thus, not allowed to overlap but can be nested.
An i-forest is \emph{restricted} if each of its boxes contains only renormalization parts.
To each restricted i-forest $\mathscr{F}$ one associates again an amplitude, namely
\begin{equation}\label{Restricted_iforest_amplitude}
\Omega^{\mathscr{F}} \equiv
  \tilde{\prod_{\gamma \in \mathscr{F}}}
      (-t_{\gamma}) \mathcal{A}^{\mathcal{G}} \ ,
\end{equation}
where the tilde over the product sign stands for the fact that in case of nested diagrams
within the i-forest one has to apply the Taylor operators from the
innermost to the outermost diagram while for disjoint sub-diagrams
the expressions are naturally independent of the order of the Taylor
operators by virtue of \refeq{factorization}. Each i-forest corresponds to a particular diagram with counterterms. The topology of the resulting diagram consists of the original diagram $\mathcal{G}$ where the sub-diagrams $\gamma$ of the i-forest have been contracted into vertices corresponding to counterterms. The nature of the counterterms depends on the i-forest.
From now on, as a shorthand notation, a diagram where an i-forest $\mathscr{F}$ is pictured with boxes will represent directly the amplitude $\Omega^{\mathscr{F}}$. See \reffig{1loop_iForestNotation} for an example based on a one-loop diagram.

Eventually, the forest formula states that the renormalized amplitude of the
diagram $\mathcal{G}$ is given by the sum over all restricted i-forests, \textit{i.e.}\
\begin{equation}
  R^{\mathcal{G}}=\sum_{\mathscr{F} \in \mathcal{F}_{R}(\mathcal{G})}
                    \Omega^{\mathscr{F}} \ ,
\end{equation}
where $\mathcal{F}_{R}(\mathcal{G})$ denotes the set of restricted i-forests
and where it is understood that the empty i-forest (\textit{i.e.}\ without any box around a sub-diagram)
stands for the diagram $\mathcal{G}$ itself. The term with the empty i-forest corresponds to the UV divergent
diagram while all the other terms correspond to diagrams
including counterterm vertices cancelling the original UV divergences.

\section{Renormalization of in-medium diagrams}
\label{sec:RenoMB}

Theorems introduced in \refsec{sec:BasicTheo} are valid regardless of the peculiar
perturbation theory at stake \textit{i.e.}\ they can be used to analyse UV divergencies
and derive sufficient counterterms whether the reference state is $\ket{0}$ or
$\ket{\Phi^\mathrm{A}_0}$~\cite{Drissi2018}. However, as argued in \refsec{subsec:RenoAbody}, the direct use of BPHZ will generate $\mathrm{A}$-dependent counterterms that can and should be avoided.

A key component of the renormalization procedure exposed in this section
is to relate UV divergences occurring in the calculation of the approximated
in-medium Green's functions (\textit{i.e.}\ with $\ket{\Phi^\mathrm{A}_0}$ as a reference state)
to UV divergences occurring in the calculation of in-vacuum Green's functions
(\textit{i.e.}\ with $\ket{0}$ as a reference state) in a related approximation.

\subsection{Cutting procedure}

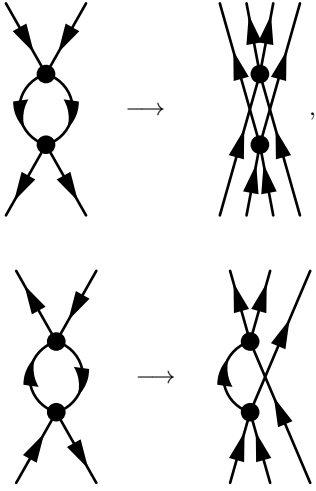
\begin{figure}
  \centering
\parbox{50pt}{
\begin{fmffile}{Figures/diags/1Loop_HoleExt}
\begin{fmfgraph*}(30,80)
\fmfstraight
\fmfbottom{i1,i2}
\fmftop{o1,o2}
\fmf{fermion}{v1,i1}
\fmf{fermion}{v1,i2}
\fmf{fermion, left=0.75}{v2,v1}
\fmf{fermion, right=0.75}{v2,v1}
\fmf{fermion}{o2,v2}
\fmf{fermion}{o1,v2}
\fmfv{d.shape=circle,d.filled=full,d.size=3thick}{v1}
\fmfv{d.shape=circle,d.filled=full,d.size=3thick}{v2}
\end{fmfgraph*}
\end{fmffile}
}
\hspace{-0.4cm}
$\longrightarrow$
\hspace{0.5cm}
\parbox{50pt}{
\begin{fmffile}{Figures/diags/1Loop_HoleExt_Transfo}
\begin{fmfgraph*}(30,80)
  \fmfstraight
  \fmfbottom{i1,i2,i3,i4}
  \fmftop{o1,o2,o3,o4}
  \fmf{fermion}{i2,v1}
  \fmf{fermion}{i3,v1}
  \fmf{fermion}{v2,o2}
  \fmf{fermion}{v2,o3}
\fmf{phantom, left=0.75}{v1,v2}
\fmf{phantom, right=0.75}{v1,v2}
\fmffreeze
\fmf{fermion}{i1,v2}
\fmf{fermion}{i4,v2}
\fmf{fermion}{v1,o1}
\fmf{fermion}{v1,o4}
\fmfv{d.shape=circle,d.filled=full,d.size=3thick}{v1}
\fmfv{d.shape=circle,d.filled=full,d.size=3thick}{v2}
\end{fmfgraph*}
\end{fmffile}
}
\hspace{-0.8cm}
,\hspace{-0.2cm} \\
\vspace{0.7cm}\hspace{0.7cm}
\parbox{50pt}{
\begin{fmffile}{Figures/diags/1Loop_HoleInt}
\begin{fmfgraph*}(30,80)
\fmfstraight
\fmfbottom{i1,i2}
\fmftop{o1,o2}
\fmf{fermion}{i1,v1}
\fmf{fermion}{v1,i2}
\fmf{fermion, left=0.75}{v1,v2}
\fmf{fermion, left=0.75}{v2,v1}
\fmf{fermion}{o2,v2}
\fmf{fermion}{v2,o1}
\fmfv{d.shape=circle,d.filled=full,d.size=3thick}{v1}
\fmfv{d.shape=circle,d.filled=full,d.size=3thick}{v2}
\end{fmfgraph*}
\end{fmffile}
}
\hspace{-0.4cm}
$\longrightarrow$
\hspace{0.5cm}
\parbox{50pt}{
\begin{fmffile}{Figures/diags/1Loop_HoleInt_Transfo}
\begin{fmfgraph*}(30,80)
\fmfstraight
\fmfbottom{i1,i2,i3}
\fmftop{o1,o2,o3}
\fmf{fermion}{i1,v1}
\fmf{fermion}{i2,v1}
\fmf{fermion, left=0.75}{v1,v2}
\fmf{phantom, left=0.75}{v2,v1}
\fmf{fermion}{v2,o2}
\fmf{fermion}{v2,o1}
\fmffreeze
\fmf{fermion}{v1,o3}
\fmf{fermion}{i3,v2}
\fmfv{d.shape=circle,d.filled=full,d.size=3thick}{v1}
\fmfv{d.shape=circle,d.filled=full,d.size=3thick}{v2}
\end{fmfgraph*}
\end{fmffile}
}
\caption{Examples of the cutting procedure applied to diagrams
$\mathcal{G}^{(\mathrm{A},k)}_2$ and resulting diagrams $\mathcal{G}^{(\mathrm{A},k+p)}_2$,
with $p=2$ (top) and $p=1$ (bottom).}
\label{ExTransfoPHDiag}
\end{figure}

Let $\mathcal{G}^{(\mathrm{A},k)}_n \in \mathcal{S}^{(\mathrm{A},k)}_{\text{MB}}$
and $\mathcal{A}^{\mathcal{G}^{(\mathrm{A},k)}_n}$ be its associated amplitude
given in \refeq{AmplitudePH}. For any finite $\mathrm{A}$, the set of states $(\vec{p}_i,\sigma_i)$ labelling unperturbed hole propagators lie in a compact space\footnote{In the limit of infinite matter at temperature $T=0$,
this remains true for a reference state associated to a compact Fermi surface.}.
Consequently, to prove the UV convergence of \refeq{AmplitudePH},
it is sufficient to prove the convergence of the sub-integral related to
the sole particle propagators. The UV behaviour
of $\mathcal{G}^{(\mathrm{A},k)}_n$ is the same as the UV behaviour of an associated
diagram made only of particle propagators.
Such diagram can be defined as the one made out of the same $n$ vertices but with
incoming particle lines $E_{in}^{+} \cup I^{-}$, outgoing particle lines $E_{out}^{+} \cup I^{-}$
and internal particle lines $I^{+}$.
As each line in the aforementioned diagram corresponds to
an unperturbed particle propagator $iG^{(\mathrm{A},1)0+}$ and possesses
$(k+p)$ outgoing and incoming external lines, where $p \equiv \# I^{-}$,
the diagram is denoted in the following as $\mathcal{G}^{(\mathrm{A},k+p)}_{n}$.
Diagrammatically, $\mathcal{G}^{(\mathrm{A},k+p)}_{n}$ is obtained from the original diagram
by cutting all internal unperturbed hole lines in $\mathcal{G}^{(\mathrm{A},k)}_{n}$
and replacing each of them by an incoming and an outgoing external unperturbed particle propagator.
Examples of the cutting procedure are displayed in \reffig{ExTransfoPHDiag}.

Consequently, the cutting procedure recasts the analysis of the UV behaviour of
a diagram contributing to the in-medium $k$-body Green's function $i^k G^{(\mathrm{A},k)}$
into the analysis of a diagram solely made of unperturbed particle
propagators contributing to the in-medium $(k+p)$-body Green's function $i^{k+p}G^{(\mathrm{A},k+p)}$.

\subsection{Relation to in-vacuum diagrams}

At this stage if BPHZ is applied to the diagrams $\mathcal{G}^{(\mathrm{A},k+p)}_n$
resulting from the cutting procedure, $\mathrm{A}$-dependent counterterms will be generated.
To avoid this, each diagram $\mathcal{G}^{(\mathrm{A},k+p)}_n$ is associated to a diagram
$\mathcal{G}^{(0,k+p)}_n$ with the same UV behaviour but that is made only of \emph{in-vacuum} propagators.

This is done by noticing that the unperturbed particle one-body Green's function
$iG^{(\mathrm{A},1)0+}$ possesses the same asymptotic coefficients $\alpha\left(S\right)$
as the in-vacuum one-body Green's function $iG^{(0,1)0}$, \textit{i.e.}\ for any sub-space
$S$ of $\{ \vec{e}_\omega, \vec{e}_{p_x}, \vec{e}_{p_y}, \vec{e}_{p_z}\}$
\begin{equation}\label{EqualAsymptotCoeff}
  \alpha^+\left(S\right) = \alpha^0\left(S\right) \ ,
\end{equation}
where $\alpha^+\left(S\right)$ correspond to the asymptotic coefficients of
$iG^{(\mathrm{A},1)0+}_{\vec{p}\sigma}(\omega)$.
Applying the asymptotic theorem stated in \refsec{subsubsec:AsTheo}, one deduces that
\emph{any diagram made solely of unperturbed particle propagators
is UV-convergent if and only if the same diagram made of in-vacuum propagators is UV-convergent.}

One can thus focus on the study of the diagram $\mathcal{G}^{(0,k+p)}_n$
associated to $\mathcal{G}^{(\mathrm{A},k+p)}_n$. As detailed in \refsec{subsec:BPHZ}, the UV divergences are canceled out by additional diagrams, containing
counterterms, generated by the BPHZ procedure.
The renormalized amplitude of $\mathcal{G}^{(0,k+p)}_n$ reads as
\begin{equation}
  R^{\mathcal{G}^{(0,k+p)}_n}
      = \sum_{\mathscr{F} \in \mathcal{F}_{R}(\mathcal{G}^{(0,k+p)}_n)} \Omega^{\mathscr{F}} \ .
\end{equation}
In order to transport the cancellation of UV divergences back to $\mathcal{G}^{(\mathrm{A},k+p)}_n$, one introduces the amplitude
\begin{equation}\label{ModifiedBPHZ}
  R^{\mathcal{G}^{(\mathrm{A},k+p)}_n}
      \equiv
        \sum_{\mathscr{F} \in \mathcal{F}_{R}(\mathcal{G}^{(\mathrm{A},k+p)}_n)}
            \Omega^{(\mathrm{A})\mathscr{F}} \ ,
\end{equation}
where $\Omega^{(\mathrm{A})\mathscr{F}}$ denotes the amplitude associated to the same diagram as the one originating from
$\Omega^{\mathscr{F}}$ except that lines denoting in-vacuum propagators are replaced by unperturbed particle propagators.
Therefore, the amplitudes $\Omega^{(\mathrm{A})\mathscr{F}}$ contains $k$-body counterterms as in $\Omega^{\mathscr{F}}$ which are, by construction, the same for all $\mathrm{A}\geq k$.

\subsection{General procedure}\label{subsec:GeneralProcedure}

The procedure to derive UV-finite $k$-body Green's functions computed with respect to
$\ket{\Phi^\mathrm{A}_0}$ on the basis of a many-body approximation defined by a truncated set of particle-hole diagrams
$\mathcal{S}^{(\mathrm{A},k)}_{\text{MB}}$  is now recapitulated.
For any diagram $\mathcal{G}^{(\mathrm{A},k)} \in \mathcal{S}^{(\mathrm{A},k)}_{\text{MB}}$ with $p$ internal hole lines
\begin{enumerate}
  \item Apply the cutting procedure to $\mathcal{G}^{(\mathrm{A},k)}$, \textit{i.e.}\ cut internal hole lines
        into external particle propagators and replace external hole lines by particle propagators.
        From this first step the associated diagram $\mathcal{G}^{(\mathrm{A},k+p)}$ is obtained.
  \item Replace in $\mathcal{G}^{(\mathrm{A},k+p)}$ all particle propagators by in-vacuum propagators, thus generating
        the diagram $\mathcal{G}^{(0,k+p)}$ contributing to the $(k+p)$-body
        Green's function with respect to $\ket{0}$.
\end{enumerate}
The set of diagrams thus obtained is denoted as $\mathcal{S}^{(0,k+p)}_{\text{MB}}$.
\begin{enumerate}[resume]
  \item The renormalization of LECs can be carried out as usual on  $\mathcal{G}^{(0,k+p)}$, leading to the introduction of an additional set of diagrams with counterterms
        denoted as $\mathcal{S}^{(0,k+p)}_{\text{MB,ct}}$.
        The LECs are typically fixed by matching (a subset of) the $(k+p)$-body Green's functions to observables in
        their approximations defined by
        $\mathcal{S}^{(0,k+p)}_{\text{MB}} \cup \mathcal{S}^{(0,k+p)}_{\text{MB,ct}}$.
\end{enumerate}
Then, for any diagram $\mathcal{G}^{(0,k+p)}_{\text{ct}} \in \mathcal{S}^{(0,k+p)}_{\text{MB,ct}}$,
\begin{enumerate}[resume]
  \item Replace each in-vacuum propagator with a particle propagator. This generates the diagram
        $\mathcal{G}^{(\mathrm{A},k+p)}_{\text{ct}}$.
  \item External lines obtained via the cut (replacement) of internal hole lines are closed (replaced)
        by a hole line. This leads to the diagram $\mathcal{G}^{(\mathrm{A},k)}_{\text{ct}}$.
\end{enumerate}
The set of all diagrams $\mathcal{G}^{(\mathrm{A},k)}_{\text{ct}}$ thus obtained is denoted as
$\mathcal{S}^{(\mathrm{A},k)}_{\text{MB,ct}}$.
\begin{enumerate}[resume]
  \item The approximated UV-finite $k$-body Green's function eventually reads as \refeq{Approx_kGF_PH_CT}, \textit{i.e.}
        \begin{align*}
            &i^k G^{(\mathrm{A},k)\text{RMB}}_{\substack{\mu_1 \dots \mu_k \\ \nu_1 \dots \nu_k}}
            (\omega_{\mu_1} \dots \omega_{\mu_k} ,\omega_{\nu_1} \dots \omega_{\nu_k}) \equiv
              \\
              &
                \sum_{\mathcal{G}^{(\mathrm{A},k)} \in
                          \mathcal{S}^{(\mathrm{A},k)}_{\text{MB}} \cup
                          \mathcal{S}^{(\mathrm{A},k)}_{\text{MB,ct}}
                     }
                 \mathcal{A}^{\mathcal{G}^{(\mathrm{A},k)}}_{\substack{\mu_1 \dots \mu_k \\ \nu_1 \dots \nu_k}}
                 (\omega_{\mu_1} \dots \omega_{\mu_k} ,\omega_{\nu_1} \dots \omega_{\nu_k})
           \ .
       \end{align*}
\end{enumerate}

\subsection{Discussion}

Section~\ref{sec:RenoMB} introduced a general procedure to transpose a renormalization scheme formulated for a perturbative expansion around the particle vacuum $\ket{0}$ to a perturbative expansion using an $\mathrm{A}$-body reference state $\ket{\Phi^\mathrm{A}_0}$ whilst keeping the same partitioning of $H$ \textit{i.e.}\ taking $H_0$ as the kinetic energy.
Given an approximation to the in-medium $k$-body Green's function, $\mathrm{A}$-independent counterterms can be fixed by matching in-vacuum $(k+p)$-body Green's functions $i^{k+p}G^{(0,k+p)}$ to appropriate observables instead of in-medium $k$-body ones $i^{k}G^{(\mathrm{A},k)}$. Depending on the approximation, the index $p$ may span a large range of values, \textit{e.g.}\ one may consider diagrams containing an arbitrary large number of hole lines $p$. A key practical point relates thus to which subset of those $(k+p)$-body Green's functions $i^{k+p}G^{(0,k+p)}$ actually need to be used to fix the LECs. Indeed, if the range of $(k+p)$-body sectors effectively needed  to renormalize a given set of many-body diagrams is not small, the procedure will be laborious. In practice, however, the BPHZ procedure stipulates that only 1PI sub-diagrams that are superficially divergent, the so-called renormalization parts, must be considered. This hopefully limits the possible topology of the renormalization parts and of the associated counterterms such that the set of in-vacuum Green's functions that needs to be eventually considered to achieve renormalization remains very limited.

The hope is thus that the set of identified counterterms is gentle. For pure neutron systems, where the three-body contact term entering $H^{\text{LO}}_{\spi}$ is inactive, renormalization was shown to be achievable with only $\delta C_0(\Lambda)$ counterterms on the basis of \emph{exact} (no approximation) vacuum two- and three-body Green's functions~\cite{Bedaque2000}. If both neutrons and protons are present, the $D_0(\Lambda)$ three-body contact term is necessary to achieve renormalization. Numerical calculations seem to indicate that this conclusion extends to the vacuum four-body Green's function \cite{Platter2004}.

As a matter of fact, the set of counterterms needed to handle a given many-body approximation is mainly driven by two important features, \textit{i.e.}\ (i) the topology of the diagrams controlling their ultraviolet character and (ii) the degeneracy factor $g$ of the interacting fermions, \textit{e.g.}\ $g=2$ for spin one-half neutrons and $g=4$ if protons are added. While both considerations are of different nature, the second one must not be overlooked in practice given that Pauli blocking can forbid certain topologies and their otherwise divergent character, thus drastically limiting the number of potentially needed counterterms compared to the case of an arbitrary $g$. As discussed in details for the dilute fermi gas in Ref.~\cite{Wellenhofer:2018dwh}, a sole\footnote{The focus is here on starting from a bare $C_0$ term appropriate to accounting for the S-wave scattering length $a_0$. The $C_2$ term necessary to describe the effective range $r_0$ requires new counterterms at each order.} $\delta C_0(\Lambda)$ counterterm is needed to achieve renormalization up to 4$^{\text{th}}$ order in perturbation theory for $g=2$ whereas an additional $D_0(\Lambda)$
three-body contact term is necessary for $g=4$. Going to higher orders, diagrams with a larger number of particle lines propagating simultaneously are allowed, at least if the particle number $A$ is large enough\footnote{If the number of particles is not at least equal to the number of simultaneously propagating particle lines, the diagram is zero, which is why few-body sectors are a priori better protected, \textit{e.g.}\ the set of diagrams providing an exact treatment of two-body systems is limited to ladder diagrams with only two particle lines propagating simultaneously. As A increases, diagrams with richer and richer topologies arise, especially as one goes to higher and higher orders in perturbation theory.} potentially leading to renormalization parts with more external legs and thus requiring counterterms of higher ranks. Given that $k$-body contact terms are Pauli blocked beyond $k=2$ ($k=4$) for $g=2$ ($g=4$), this situation would require derivative terms. While being speculative, this situation cannot be excluded at this point as there exists no general proof, to our knowledge, that it is not the case.

Starting from a bare $C_0$ and \emph{conjecturing} that $\delta C_0(\Lambda)$ counterterms are sufficient to obtain UV-finite $\mathcal{G}^{(0,k)}_{n}$ in pure neutron systems for any value of $k$ and $n$, this property would remain true for any $\mathcal{G}^{(\mathrm{A},k)}_n$ thanks to the procedure discussed in \refsec{subsec:GeneralProcedure}. In terms of the subtracting operator $t_\gamma$, the conjecture consists in assuming that there exists an operator with $t_\gamma \neq 0$ only for two-body diagrams $\gamma$. This would mean that the set of renormalization parts could be restricted to two-body sub-diagrams such that the renormalization in the two-body sector would be sufficient to ensure it for any perturbative approximation and any particle number. As mentioned above, and to the best of our knowledge, this conjecture remains however unproven. Proving it would be an important step forward in the understanding of $\spi$EFT and its application to all $\mathrm{A}$-body sectors.


\section{Application and extension}
\label{sec:AppAndExt}

In this section, the procedure described in \refsec{subsec:GeneralProcedure} is applied to the calculation of the in-medium one-body propagator within the random phase approximation (RPA). The well-celebrated RPA truncation scheme acts as an example of practical interest for many-body calculations.
Then, an extension of the general procedure is discussed in connection with the use of a more general partitioning of $H$, namely the Hartree-Fock (HF) partitioning, to define the perturbation theory and the associated one-body unperturbed propagators.

\subsection{Random phase approximation}
\label{subsec:RPA}

Historically, the RPA was first introduced in Ref.~\cite{Bohm1953} as a way to deal with collective phenomena
such as charge screening effect in the electronic gas. Later, it was reformulated in Refs.~\cite{Gell-Mann1957,Hubbard1957}
as a particular resummation of perturbation theory diagrams, namely forward and backward particle-hole excitations.
Regarding nuclear systems, RPA and its extensions play an important role to tackle collective excitations~\cite{Czerski1986}.

\begin{figure}
\centering
\parbox{50pt}{\begin{fmffile}{Figures/diags/1PISelf_RPA}
   \begin{fmfgraph*}(30,80)
   \fmfset{arrow_len}{2.8mm}
   \fmfset{arrow_ang}{20}
   \fmfbottom{b1} \fmftop{t1}
   \fmf{fermion}{b1,v1}
   \fmf{fermion}{v1,t1}
   \fmfv{d.sh=circle,d.f=empty, d.si=1w, b=(1,,1,,1), label=$\Sigma^{\text{RPA}}$, label.dist=0}{v1}
  \end{fmfgraph*}
\end{fmffile}}
\hspace{-0.7cm}
=
\hspace{-0.4cm}
\parbox{50pt}{\begin{fmffile}{Figures/diags/1stOrder1PISelf_RPA}
  \begin{fmfgraph}(30,80)
    \fmfkeep{1stOrder1PISelf_RPA}
  \fmfbottom{i}
  \fmftop{o}
  \fmf{fermion}{i,v1}
  \fmf{fermion}{v1,o}
  \fmf{fermion, tension=1}{v1,v1}
  \fmfv{d.shape=circle,d.filled=full,d.size=2thick}{v1}
  \end{fmfgraph}
\end{fmffile}}
\hspace{-0.1cm}
+
\hspace{0.1cm}
\parbox{50pt}{\begin{fmffile}{Figures/diags/2ndOrder1PISelf_RPA}
  \begin{fmfgraph}(30,80)
    \fmfkeep{2ndOrder1PISelf_RPA}
  \fmfbottom{i}
  \fmftop{o}
  \fmf{fermion, tag=1}{i,v1}
  \fmf{fermion, tag=3}{v1,v2}
  \fmf{fermion, tag=5}{v2,o}
  \fmffreeze
  \fmf{fermion, right, tag=4}{v1,v2}
  \fmffreeze
  \fmf{fermion, right, tag=2}{v2,v1}
  \fmfv{d.shape=circle,d.filled=full,d.size=2thick}{v1}
  \fmfv{d.shape=circle,d.filled=full,d.size=2thick}{v2}
  \end{fmfgraph}
\end{fmffile}}
\hspace{-0.6cm}
+
\hspace{-0.4cm}
\parbox{50pt}{\begin{fmffile}{Figures/diags/3rdOrder1PISelf_RPA_1}
  \begin{fmfgraph}(30,80)
    \fmfkeep{3rdOrder1PISelf_RPA_1}
  \fmfbottom{i}
  \fmftop{o}
  \fmfright{r}
  \fmf{fermion, tension=4}{i,v1}
  \fmf{fermion}{v1,v3}
  \fmf{fermion, tension=4}{v3,o}
  \fmffreeze
  \fmf{fermion, right=0.25}{v1,v2}
  \fmf{fermion, right=0.25}{v2,v3}
  \fmf{phantom, tension=20}{v2,r}
  \fmffreeze
  \fmf{fermion, right=0.25}{v3,v2}
  \fmf{fermion, right=0.25}{v2,v1}
  \fmfv{d.shape=circle,d.filled=full,d.size=2thick}{v1}
  \fmfv{d.shape=circle,d.filled=full,d.size=2thick}{v2}
  \fmfv{d.shape=circle,d.filled=full,d.size=2thick}{v3}
  \end{fmfgraph}
\end{fmffile}}
\hspace{-0.6cm}
+
\hspace{-0.4cm}
\parbox{50pt}{\begin{fmffile}{Figures/diags/4thOrder1PISelf_RPA}
  \begin{fmfgraph}(30,80)
    \fmfkeep{4thOrder1PISelf_RPA}
    \fmfbottom{i}
    \fmftop{o}
    \fmfright{r}
    \fmf{fermion, tension=10}{i,v1}
    \fmf{fermion}{v1,v4}
    \fmf{fermion, tension=10}{v4,o}
    \fmffreeze
    \fmf{fermion, right=0.25, tension=13}{v1,v2}
    \fmf{fermion, right=0.25, tension=0.1}{v2,v3}
    \fmf{fermion, right=0.25, tension=13}{v3,v4}
    \fmf{phantom, tension=30}{v2,r}
    \fmf{phantom, tension=30}{v3,r}
    \fmffreeze
    \fmf{fermion, right=0.25}{v4,v3}
    \fmf{fermion, right=0.25}{v3,v2}
    \fmf{fermion, right=0.25}{v2,v1}
    \fmfv{d.shape=circle,d.filled=full,d.size=2thick}{v1}
    \fmfv{d.shape=circle,d.filled=full,d.size=2thick}{v2}
    \fmfv{d.shape=circle,d.filled=full,d.size=2thick}{v3}
    \fmfv{d.shape=circle,d.filled=full,d.size=2thick}{v4}
  \end{fmfgraph}
\end{fmffile}}
\hspace{-0.6cm}
+
\dots

\caption{Examples of 1PI time-unordered diagrams contributing to the one-body Green's function
in the RPA approximation. Oriented lines denote here the unperturbed propagator \eqref{UnpertProp}.}
\label{FigSelfEnergy1PI_RPA}
\end{figure}
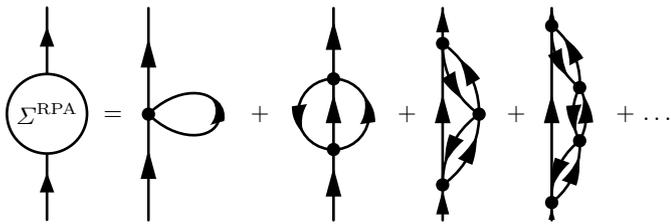

Here the Hamiltonian is partitioned as in \refeqs{kineticPartH} and the reference state is chosen to be
$| \Phi^\mathrm{A}_0 \rangle$ as defined in \refeq{refStateA}. The RPA is formulated as an approximation
to the self-energy in the Dyson equation, namely
\begin{multline}\label{DysonEqRPA}
  G^{(\mathrm{A},1)\text{RPA}}_{\mu\nu}(\omega) =
    G^{(\mathrm{A},1)0}_{\mu\nu}(\omega) \\
      + \sum_{\lambda_1 \lambda_2}
        G^{(\mathrm{A},1)\text{RPA}}_{\mu\lambda_1}(\omega)
        \ \Sigma^{(\mathrm{A})\text{RPA}}_{\lambda_1\lambda_2}(\omega)
        \ G^{(\mathrm{A},1)0}_{\lambda_2\nu}(\omega) \ ,
\end{multline}
where $G^{(\mathrm{A},1)\text{RPA}}_{\mu\nu}(\omega)$ and
$\Sigma^{(\mathrm{A})\text{RPA}}_{\lambda_1\lambda_2}(\omega)$ denote respectively
the one-body propagator and the self-energy in the RPA approximation, and the
unperturbed propagator reads as
\begin{equation}\label{UnpertProp}
  iG^{(\mathrm{A},1)0}_{\mu\nu}(\omega) =
    iG^{(\mathrm{A},1)0+}_{\mu\nu}(\omega)
    + iG^{(\mathrm{A},1)0-}_{\mu\nu}(\omega) \ .
\end{equation}
The 1PI time-unordered Feynman diagrams contributing to the self-energy
in this approximation consist of the so-called ring diagrams.
Examples of first, second, third and fourth order (in terms of
number of vertices) contributions  to the self-energy are displayed in~\reffig{FigSelfEnergy1PI_RPA}.
Once the counterterms are correctly taken into account for the 1PI part of the one-body Green's function,
no additional UV-divergences appear in the full (1PR) one-body Green's function. Thus, in the following,
we focus only on 1PI diagrams.

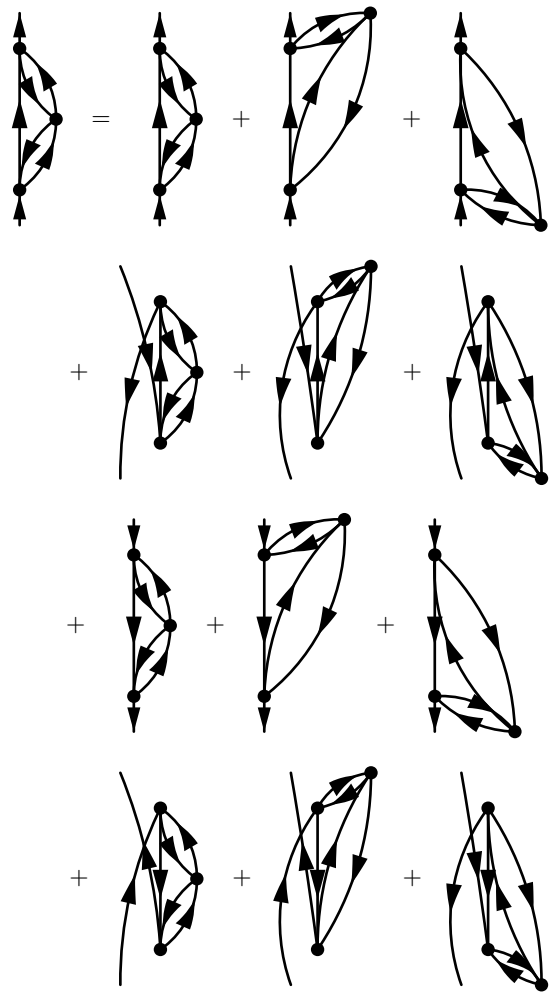
\begin{figure}
\centering
\hspace{-0.3cm}
\parbox{50pt}{\begin{fmffile}{Figures/diags/3rdOrder1PISelf_RPA_1_TU}
  \begin{fmfgraph}(30,80)
    \fmfkeep{3rdOrder1PISelf_RPA_1_TU}
  \fmfstraight
  \fmfbottom{i}
  \fmftop{o}
  \fmfright{r}
  \fmf{fermion, tension=4}{i,v1}
  \fmf{fermion}{v1,v3}
  \fmf{fermion, tension=4}{v3,o}
  \fmffreeze
  \fmf{fermion, right=0.25}{v1,v2}
  \fmf{fermion, right=0.25}{v2,v3}
  \fmf{phantom, tension=20}{v2,r}
  \fmffreeze
  \fmf{fermion, right=0.25}{v3,v2}
  \fmf{fermion, right=0.25}{v2,v1}
  \fmfv{d.shape=circle,d.filled=full,d.size=2thick}{v1}
  \fmfv{d.shape=circle,d.filled=full,d.size=2thick}{v2}
  \fmfv{d.shape=circle,d.filled=full,d.size=2thick}{v3}
  \end{fmfgraph}
\end{fmffile}}
\hspace{-0.5cm}
=
\hspace{-0.1cm}
\parbox{50pt}{\begin{fmffile}{Figures/diags/3rdOrder1PISelf_RPA_1_TO}
  \begin{fmfgraph}(30,80)
    \fmfkeep{3rdOrder1PISelf_RPA_1_TO}
  \fmfstraight
  \fmfbottom{i}
  \fmftop{o}
  \fmfright{r}
  \fmf{fermion, tension=4}{i,v1}
  \fmf{fermion}{v1,v3}
  \fmf{fermion, tension=4}{v3,o}
  \fmffreeze
  \fmf{fermion, right=0.25}{v1,v2}
  \fmf{fermion, right=0.25}{v2,v3}
  \fmf{phantom, tension=20}{v2,r}
  \fmffreeze
  \fmf{fermion, right=0.25}{v3,v2}
  \fmf{fermion, right=0.25}{v2,v1}
  \fmfv{d.shape=circle,d.filled=full,d.size=2thick}{v1}
  \fmfv{d.shape=circle,d.filled=full,d.size=2thick}{v2}
  \fmfv{d.shape=circle,d.filled=full,d.size=2thick}{v3}
  \end{fmfgraph}
\end{fmffile}}
\hspace{-0.5cm}
+
\hspace{0.3cm}
\parbox{50pt}{\begin{fmffile}{Figures/diags/3rdOrder1PISelf_RPA_2_TO}
  \begin{fmfgraph}(30,80)
    \fmfkeep{3rdOrder1PISelf_RPA_2_TO}
  \fmfstraight
  \fmfbottom{i1,i2}
  \fmftop{o1,o2}
  \fmf{fermion, tension=4}{i1,v1}
  \fmf{fermion}{v1,v3}
  \fmf{fermion, tension=4}{v3,o1}
  \fmffreeze
  \fmf{fermion, left=0.25}{v3,o2}
  \fmf{fermion, left=0.20}{o2,v3}
  \fmffreeze
  \fmf{fermion, left=0.20}{v1,o2}
  \fmf{fermion, left=0.25}{o2,v1}
  \fmfv{d.shape=circle,d.filled=full,d.size=2thick}{v1}
  \fmfv{d.shape=circle,d.filled=full,d.size=2thick}{o2}
  \fmfv{d.shape=circle,d.filled=full,d.size=2thick}{v3}
  \end{fmfgraph}
\end{fmffile}}
\hspace{-0.5cm}
+
\hspace{0.3cm}
\parbox{50pt}{\begin{fmffile}{Figures/diags/3rdOrder1PISelf_RPA_3_TO}
  \begin{fmfgraph}(30,80)
    \fmfkeep{3rdOrder1PISelf_RPA_3_TO}
  \fmfstraight
  \fmfbottom{i1,i2}
  \fmftop{o1,o2}
  \fmf{fermion, tension=4}{i1,v1}
  \fmf{fermion}{v1,v3}
  \fmf{fermion, tension=4}{v3,o1}
  \fmffreeze
  \fmf{fermion, left=0.25}{v1,i2}
  \fmf{fermion, left=0.20}{i2,v1}
  \fmffreeze
  \fmf{fermion, left=0.20}{v3,i2}
  \fmf{fermion, left=0.25}{i2,v3}
  \fmfv{d.shape=circle,d.filled=full,d.size=2thick}{v1}
  \fmfv{d.shape=circle,d.filled=full,d.size=2thick}{i2}
  \fmfv{d.shape=circle,d.filled=full,d.size=2thick}{v3}
  \end{fmfgraph}
\end{fmffile}} \\
\vspace{0.5cm}\hspace{1cm}
+
\hspace{0.2cm}
\parbox{50pt}{\begin{fmffile}{Figures/diags/3rdOrder1PISelf_RPA_1_TO_Crossed_HH}
  \begin{fmfgraph}(30,80)
    \fmfkeep{3rdOrder1PISelf_RPA_1_TO_Crossed_HH}
  \fmfstraight
  \fmfbottom{i1,i2,i3}
  \fmftop{o1,o2,o3}
  \fmfright{r}
  \fmf{phantom, tension=4}{i2,v1}
  \fmf{fermion}{v1,v3}
  \fmf{phantom, tension=4}{v3,o2}
  \fmffreeze
  \fmf{fermion, right=0.25}{v1,v2}
  \fmf{fermion, right=0.25}{v2,v3}
  \fmf{phantom, tension=20}{v2,r}
  \fmffreeze
  \fmf{fermion, right=0.25}{v3,v2}
  \fmf{fermion, right=0.25}{v2,v1}
  \fmffreeze
  \fmf{fermion, left=0.08}{o1,v1}
  \fmf{fermion, right=0.115}{v3,i1}
  \fmfv{d.shape=circle,d.filled=full,d.size=2thick}{v1}
  \fmfv{d.shape=circle,d.filled=full,d.size=2thick}{v2}
  \fmfv{d.shape=circle,d.filled=full,d.size=2thick}{v3}
  \end{fmfgraph}
\end{fmffile}}
\hspace{-0.5cm}
+
\hspace{0.3cm}
\parbox{50pt}{\begin{fmffile}{Figures/diags/3rdOrder1PISelf_RPA_2_TO_Crossed_HH}
  \begin{fmfgraph}(30,80)
    \fmfkeep{3rdOrder1PISelf_RPA_2_TO_Crossed_HH}
  \fmfstraight
  \fmfbottom{i1,i2,i3,i4}
  \fmftop{o1,o2,o3,o4}
  \fmf{phantom, tension=4}{i2,v1}
  \fmf{fermion}{v1,v3}
  \fmf{phantom, tension=4}{v3,o2}
  \fmffreeze
  \fmf{fermion, left=0.25}{v3,o4}
  \fmf{fermion, left=0.20}{o4,v3}
  \fmffreeze
  \fmf{fermion, left=0.15}{v1,o4}
  \fmf{fermion, left=0.15}{o4,v1}
  \fmffreeze
  \fmf{fermion, left=0.01}{o1,v1}
  \fmf{fermion, right=0.25}{v3,i1}
  \fmfv{d.shape=circle,d.filled=full,d.size=2thick}{v1}
  \fmfv{d.shape=circle,d.filled=full,d.size=2thick}{o4}
  \fmfv{d.shape=circle,d.filled=full,d.size=2thick}{v3}
  \end{fmfgraph}
\end{fmffile}}
\hspace{-0.5cm}
+
\hspace{0.3cm}
\parbox{50pt}{\begin{fmffile}{Figures/diags/3rdOrder1PISelf_RPA_3_TO_Crossed_HH}
  \begin{fmfgraph}(30,80)
    \fmfkeep{3rdOrder1PISelf_RPA_3_TO_Crossed_HH}
  \fmfstraight
  \fmfbottom{i1,i2,i3,i4}
  \fmftop{o1,o2,o3,o4}
  \fmf{phantom, tension=4}{i2,v1}
  \fmf{fermion}{v1,v3}
  \fmf{phantom, tension=4}{v3,o2}
  \fmffreeze
  \fmf{fermion, left=0.25}{v1,i4}
  \fmf{fermion, left=0.20}{i4,v1}
  \fmffreeze
  \fmf{fermion, left=0.15}{v3,i4}
  \fmf{fermion, left=0.15}{i4,v3}
  \fmffreeze
  \fmf{fermion, left=0.01}{o1,v1}
  \fmf{fermion, right=0.25}{v3,i1}
  \fmfv{d.shape=circle,d.filled=full,d.size=2thick}{v1}
  \fmfv{d.shape=circle,d.filled=full,d.size=2thick}{i4}
  \fmfv{d.shape=circle,d.filled=full,d.size=2thick}{v3}
  \end{fmfgraph}
\end{fmffile}}
\hspace{-0.5cm}
\\
\vspace{0.5cm}\hspace{0.5cm}
+
\parbox{50pt}{\begin{fmffile}{Figures/diags/3rdOrder1PISelf_RPA_1_TO_HH}
  \begin{fmfgraph}(30,80)
    \fmfkeep{3rdOrder1PISelf_RPA_1_TO_HH}
  \fmfstraight
  \fmfbottom{i}
  \fmftop{o}
  \fmfright{r}
  \fmf{fermion, tension=4}{v1,i}
  \fmf{fermion}{v3,v1}
  \fmf{fermion, tension=4}{o,v3}
  \fmffreeze
  \fmf{fermion, right=0.25}{v1,v2}
  \fmf{fermion, right=0.25}{v2,v3}
  \fmf{phantom, tension=20}{v2,r}
  \fmffreeze
  \fmf{fermion, right=0.25}{v3,v2}
  \fmf{fermion, right=0.25}{v2,v1}
  \fmfv{d.shape=circle,d.filled=full,d.size=2thick}{v1}
  \fmfv{d.shape=circle,d.filled=full,d.size=2thick}{v2}
  \fmfv{d.shape=circle,d.filled=full,d.size=2thick}{v3}
  \end{fmfgraph}
\end{fmffile}}
\hspace{-0.5cm}
+
\hspace{0.3cm}
\parbox{50pt}{\begin{fmffile}{Figures/diags/3rdOrder1PISelf_RPA_2_TO_HH}
  \begin{fmfgraph}(30,80)
    \fmfkeep{3rdOrder1PISelf_RPA_2_TO_HH}
  \fmfstraight
  \fmfbottom{i1,i2}
  \fmftop{o1,o2}
  \fmf{fermion, tension=4}{v1,i1}
  \fmf{fermion}{v3,v1}
  \fmf{fermion, tension=4}{o1,v3}
  \fmffreeze
  \fmf{fermion, left=0.25}{v3,o2}
  \fmf{fermion, left=0.20}{o2,v3}
  \fmffreeze
  \fmf{fermion, left=0.20}{v1,o2}
  \fmf{fermion, left=0.25}{o2,v1}
  \fmfv{d.shape=circle,d.filled=full,d.size=2thick}{v1}
  \fmfv{d.shape=circle,d.filled=full,d.size=2thick}{o2}
  \fmfv{d.shape=circle,d.filled=full,d.size=2thick}{v3}
  \end{fmfgraph}
\end{fmffile}}
\hspace{-0.5cm}
+
\hspace{0.3cm}
\parbox{50pt}{\begin{fmffile}{Figures/diags/3rdOrder1PISelf_RPA_3_TO_HH}
  \begin{fmfgraph}(30,80)
    \fmfkeep{3rdOrder1PISelf_RPA_3_TO_HH}
  \fmfstraight
  \fmfbottom{i1,i2}
  \fmftop{o1,o2}
  \fmf{fermion, tension=4}{v1,i1}
  \fmf{fermion}{v3,v1}
  \fmf{fermion, tension=4}{o1,v3}
  \fmffreeze
  \fmf{fermion, left=0.25}{v1,i2}
  \fmf{fermion, left=0.20}{i2,v1}
  \fmffreeze
  \fmf{fermion, left=0.20}{v3,i2}
  \fmf{fermion, left=0.25}{i2,v3}
  \fmfv{d.shape=circle,d.filled=full,d.size=2thick}{v1}
  \fmfv{d.shape=circle,d.filled=full,d.size=2thick}{i2}
  \fmfv{d.shape=circle,d.filled=full,d.size=2thick}{v3}
  \end{fmfgraph}
\end{fmffile}} \\
\vspace{0.5cm}\hspace{0.5cm}
+
\hspace{0.2cm}
\parbox{50pt}{\begin{fmffile}{Figures/diags/3rdOrder1PISelf_RPA_1_TO_Crossed_PP}
  \begin{fmfgraph}(30,80)
    \fmfkeep{3rdOrder1PISelf_RPA_1_TO_Crossed_PP}
  \fmfstraight
  \fmfbottom{i1,i2,i3}
  \fmftop{o1,o2,o3}
  \fmfright{r}
  \fmf{phantom, tension=4}{v1,i2}
  \fmf{fermion}{v3,v1}
  \fmf{phantom, tension=4}{o2,v3}
  \fmffreeze
  \fmf{fermion, right=0.25}{v1,v2}
  \fmf{fermion, right=0.25}{v2,v3}
  \fmf{phantom, tension=20}{v2,r}
  \fmffreeze
  \fmf{fermion, right=0.25}{v3,v2}
  \fmf{fermion, right=0.25}{v2,v1}
  \fmffreeze
  \fmffreeze
  \fmf{fermion, right=0.08}{v1,o1}
  \fmf{fermion, left=0.115}{i1,v3}
  \fmfv{d.shape=circle,d.filled=full,d.size=2thick}{v1}
  \fmfv{d.shape=circle,d.filled=full,d.size=2thick}{v2}
  \fmfv{d.shape=circle,d.filled=full,d.size=2thick}{v3}
  \end{fmfgraph}
\end{fmffile}}
\hspace{-0.5cm}
+
\hspace{0.3cm}
\parbox{50pt}{\begin{fmffile}{Figures/diags/3rdOrder1PISelf_RPA_2_TO_Crossed_PP}
  \begin{fmfgraph}(30,80)
    \fmfkeep{3rdOrder1PISelf_RPA_2_TO_Crossed_PP}
  \fmfstraight
  \fmfbottom{i1,i2,i3,i4}
  \fmftop{o1,o2,o3,o4}
  \fmf{phantom, tension=4}{v1,i2}
  \fmf{fermion}{v3,v1}
  \fmf{phantom, tension=4}{o2,v3}
  \fmffreeze
  \fmf{fermion, left=0.25}{v3,o4}
  \fmf{fermion, left=0.20}{o4,v3}
  \fmffreeze
  \fmf{fermion, left=0.15}{v1,o4}
  \fmf{fermion, left=0.15}{o4,v1}
  \fmffreeze
  \fmf{fermion, right=0.01}{v1,o1}
  \fmf{fermion, left=0.25}{i1,v3}
  \fmfv{d.shape=circle,d.filled=full,d.size=2thick}{v1}
  \fmfv{d.shape=circle,d.filled=full,d.size=2thick}{o4}
  \fmfv{d.shape=circle,d.filled=full,d.size=2thick}{v3}
  \end{fmfgraph}
\end{fmffile}}
\hspace{-0.5cm}
+
\hspace{0.3cm}
\parbox{50pt}{\begin{fmffile}{Figures/diags/3rdOrder1PISelf_RPA_3_TO_Crossed_PP}
  \begin{fmfgraph}(30,80)
    \fmfkeep{3rdOrder1PISelf_RPA_3_TO_Crossed_PP}
  \fmfstraight
  \fmfbottom{i1,i2,i3,i4}
  \fmftop{o1,o2,o3,o4}
  \fmf{phantom, tension=4}{v1,i2}
  \fmf{fermion}{v3,v1}
  \fmf{phantom, tension=4}{o2,v3}
  \fmffreeze
  \fmf{fermion, left=0.25}{v1,i4}
  \fmf{fermion, left=0.20}{i4,v1}
  \fmffreeze
  \fmf{fermion, left=0.15}{v3,i4}
  \fmf{fermion, left=0.15}{i4,v3}
  \fmffreeze
  \fmf{fermion, left=0.01}{o1,v1}
  \fmf{fermion, right=0.25}{v3,i1}
  \fmfv{d.shape=circle,d.filled=full,d.size=2thick}{v1}
  \fmfv{d.shape=circle,d.filled=full,d.size=2thick}{i4}
  \fmfv{d.shape=circle,d.filled=full,d.size=2thick}{v3}
  \end{fmfgraph}
\end{fmffile}}
\hspace{-0.5cm}

\caption{Example of decomposition of a time-unordered diagram (left-hand side) into a sum of time-ordered diagrams (right-hand side).
In a time-unordered diagram, any line refers to the complete propagator $iG^{(\mathrm{A},1)0}$.
In a time-ordered diagram, any ascending (descending) line refers to the unperturbed particle (hole)
propagator $iG^{(\mathrm{A},1)0+}$ ($iG^{(\mathrm{A},1)0-}$).
Time-ordered diagrams in the first two rows contain one loop made of unperturbed
particle propagators whereas time-ordered diagrams in the last two rows contain
zero loop made of unperturbed particle propagators. Due to conservation of momentum,
diagrams with one particle and one hole external leg vanish so that they are not represented.
For a general partitioning of the Hamiltonian, however, this would not be necessarily the case.}
\label{Fig_timeunordered_decomposed_timeordered}
\end{figure}

Each 1PI time-unordered diagram contributing to the one-body Green's function is decomposed in a sum of time-ordered
diagrams. In a time-unordered diagram, each line refers to the complete
unperturbed propagator $iG^{(\mathrm{A},1)0}$.
In a time-ordered diagram, an ascending (descending) line refers to the unperturbed particle (hole) propagator
$iG^{(\mathrm{A},1)0+}$ ($iG^{(\mathrm{A},1)0-}$).
Example of this decomposition is represented in \reffig{Fig_timeunordered_decomposed_timeordered}.
Consequently, the RPA can be recast as a truncation on the sum of particle-hole diagrams.


\begin{table}
\centering

\begin{tabular}{|c|c|c|c|c|}

\hline
 $n$ \hspace{0.15cm}&
 $\hspace{-0.05cm}\mathcal{G}^{(\mathrm{A},1)}_n \in \mathcal{S}^{(\mathrm{A},1)}_{\text{RPA}}\hspace{-0.05cm}$ &
 \hspace{0.0cm} $p$ \hspace{0.25cm}&
 $\hspace{-0.05cm}\mathcal{G}^{(0,1+p)}_n \in \mathcal{S}^{(0,1+p)}_{\text{RPA}}\hspace{-0.05cm}$&
 $\hspace{-0.1cm}D(\mathcal{G}^{(0,1+p)}_n)\hspace{-0.1cm}$ \\

\hline
2 &
\parbox{50pt}{
\begin{center}
  \begin{fmffile}{Figures/diags/2ndOrder1PI_RPA}
  \begin{fmfgraph}(30,80)
    \fmfkeep{2ndOrder1PI_RPA}
  \fmfbottom{i}
  \fmftop{o}
  \fmf{fermion, tag=1}{i,v1}
  \fmf{fermion, tag=3}{v1,v2}
  \fmf{fermion, tag=5}{v2,o}
  \fmffreeze
  \fmf{fermion, right, tag=4}{v1,v2}
  \fmffreeze
  \fmf{fermion, right, tag=2}{v2,v1}
  \fmfv{d.shape=circle,d.filled=full,d.size=2thick}{v1}
  \fmfv{d.shape=circle,d.filled=full,d.size=2thick}{v2}
  \end{fmfgraph}
  \end{fmffile}
\end{center}
}
      &
1
      &
\parbox{50pt}{
\begin{center}
  \begin{fmffile}{Figures/diags/2ndOrder1PI_RPA_Cut}
  \begin{fmfgraph}(30,80)
    \fmfkeep{2ndOrder1PI_RPA_Cut}
  \fmfbottom{i1,i2}
  \fmftop{o1,o2}
  \fmf{fermion}{i1,v1}
  \fmf{fermion}{i2,v1}
  \fmf{fermion, left=0.75}{v1,v2}
  \fmf{fermion, right=0.75}{v1,v2}
  \fmf{fermion}{v2,o1}
  \fmf{fermion}{v2,o2}
  \fmfv{d.shape=circle,d.filled=full,d.size=2thick}{v1}
  \fmfv{d.shape=circle,d.filled=full,d.size=2thick}{v2}
  \end{fmfgraph}
  \end{fmffile}
\end{center}
}
      &
1    \\
\hline
3 &
\parbox{50pt}{
\begin{center}
  \begin{fmffile}{Figures/diags/3rdOrder1PI_RPA_1}
  \begin{fmfgraph}(30,80)
    \fmfkeep{3rdOrder1PI_RPA_1}
    \fmfbottom{i}
    \fmftop{o}
    \fmfright{r}
    \fmf{fermion, tension=4}{i,v1}
    \fmf{fermion}{v1,v3}
    \fmf{fermion, tension=4}{v3,o}
    \fmffreeze
    \fmf{fermion, right=0.25}{v1,v2}
    \fmf{fermion, right=0.25}{v2,v3}
    \fmf{phantom, tension=20}{v2,r}
    \fmffreeze
    \fmf{fermion, right=0.25}{v3,v2}
    \fmf{fermion, right=0.25}{v2,v1}
    \fmfv{d.shape=circle,d.filled=full,d.size=2thick}{v1}
    \fmfv{d.shape=circle,d.filled=full,d.size=2thick}{v2}
    \fmfv{d.shape=circle,d.filled=full,d.size=2thick}{v3}
  \end{fmfgraph}
  \end{fmffile}
\end{center}
}
&
2
&
\parbox{50pt}{
\begin{center}
  \begin{fmffile}{Figures/diags/3rdOrder1PI_RPA_1_Cut}
  \begin{fmfgraph}(30,80)
    \fmfkeep{3rdOrder1PI_RPA_1_Cut}
  \fmfstraight
  \fmfbottom{i1,i2,i3}
  \fmftop{o1,o2,o3}
  \fmfright{r}
  \fmf{fermion}{i1,v1}
  \fmf{fermion}{i2,v1}
  \fmf{fermion, left=0.55}{v1,v3}
  \fmf{fermion}{v3,o1}
  \fmf{fermion}{v3,o2}
\fmffreeze
  \fmf{fermion}{i3,v2}
  \fmf{fermion, right=0.40}{v1,v2}
  \fmf{fermion, right=0.40}{v2,v3}
  \fmf{phantom, tension=1}{v2,r}
  \fmf{fermion}{v2,o3}
  \fmfv{d.shape=circle,d.filled=full,d.size=2thick}{v1}
  \fmfv{d.shape=circle,d.filled=full,d.size=2thick}{v2}
  \fmfv{d.shape=circle,d.filled=full,d.size=2thick}{v3}
  \end{fmfgraph}
  \end{fmffile}
\end{center}
}
      &
-1    \\
\hline
3 &
\parbox{50pt}{
\begin{center}
  \begin{fmffile}{Figures/diags/3rdOrder1PI_RPA_2}
  \begin{fmfgraph}(30,80)
    \fmfkeep{3rdOrder1PI_RPA_2}
    \fmfbottom{i1,i2}
    \fmftop{o1,o2}
    \fmf{fermion, tension=2}{i1,v1}
    \fmf{fermion}{v1,v3}
    \fmf{fermion, tension=2}{v3,o1}
    \fmffreeze
    \fmf{fermion, left=0.25}{v3,o2}
    \fmf{fermion, left=0.20}{o2,v3}
    \fmffreeze
    \fmf{fermion, left=0.20}{v1,o2}
    \fmf{fermion, left=0.25}{o2,v1}
    \fmfv{d.shape=circle,d.filled=full,d.size=2thick}{v1}
    \fmfv{d.shape=circle,d.filled=full,d.size=2thick}{o2}
    \fmfv{d.shape=circle,d.filled=full,d.size=2thick}{v3}
  \end{fmfgraph}
  \end{fmffile}
\end{center}
}
&
2
&
\parbox{50pt}{
\begin{center}
  \begin{fmffile}{Figures/diags/3rdOrder1PI_RPA_2_Cut}
  \begin{fmfgraph}(30,80)
    \fmfkeep{3rdOrder1PI_RPA_2_Cut}
  \fmfstraight
  \fmfbottom{i1,i2,i3}
  \fmftop{o1,o2,o3}
  \fmfright{r}
  \fmf{fermion}{i1,v1}
  \fmf{fermion}{i2,v1}
  \fmf{fermion, left=0.55}{v1,v2}
  \fmf{fermion, right=0.55}{v1,v3}
  \fmf{fermion}{v2,o1}
  \fmf{fermion}{v3,o2}
  \fmf{fermion}{v3,o3}
\fmffreeze
  \fmf{fermion}{i3,v2}
  \fmf{fermion, left=0.10}{v2,v3}
  \fmf{phantom, tension=1}{v3,r}
  \fmfv{d.shape=circle,d.filled=full,d.size=2thick}{v1}
  \fmfv{d.shape=circle,d.filled=full,d.size=2thick}{v2}
  \fmfv{d.shape=circle,d.filled=full,d.size=2thick}{v3}
  \end{fmfgraph}
  \end{fmffile}
\end{center}
}
      &
-1    \\
\hline
4 &
\parbox{50pt}{
\begin{center}
  \begin{fmffile}{Figures/diags/4thOrder1PI_RPA_1}
  \begin{fmfgraph}(30,80)
    \fmfkeep{4thOrder1PI_RPA_1}
    \fmfbottom{i}
    \fmftop{o}
    \fmfright{r}
    \fmf{fermion, tension=5}{i,v1}
    \fmf{fermion}{v1,v4}
    \fmf{fermion, tension=5}{v4,o}
    \fmffreeze
    \fmf{fermion, right=0.25, tension=13}{v1,v2}
    \fmf{fermion, right=0.25, tension=0.1}{v2,v3}
    \fmf{fermion, right=0.25, tension=13}{v3,v4}
    \fmf{phantom, tension=30}{v2,r}
    \fmf{phantom, tension=30}{v3,r}
    \fmffreeze
    \fmf{fermion, right=0.25}{v4,v3}
    \fmf{fermion, right=0.25}{v3,v2}
    \fmf{fermion, right=0.25}{v2,v1}
    \fmfv{d.shape=circle,d.filled=full,d.size=2thick}{v1}
    \fmfv{d.shape=circle,d.filled=full,d.size=2thick}{v2}
    \fmfv{d.shape=circle,d.filled=full,d.size=2thick}{v3}
    \fmfv{d.shape=circle,d.filled=full,d.size=2thick}{v4}
  \end{fmfgraph}
  \end{fmffile}
\end{center}
}
&
3
&
\parbox{50pt}{
\begin{center}
  \begin{fmffile}{Figures/diags/4thOrder1PI_RPA_1_Cut}
  \begin{fmfgraph}(30,80)
    \fmfkeep{4thOrder1PI_RPA_1_Cut}
  \fmfstraight
  \fmfbottom{i1,i2,i3,i4}
  \fmftop{o1,o2,o3,o4}
  \fmfright{r}
  \fmf{fermion, tension=5}{i1,v1}
  \fmf{fermion, tension=5}{i2,v1}
  \fmf{fermion, left=0.45, tension=4}{v1,v4}
  \fmf{fermion, tension=5}{v4,o1}
  \fmf{fermion, tension=5}{v4,o2}
  \fmffreeze
  \fmf{fermion, right=0.25, tension=18}{v1,v2}
  \fmf{fermion, left=0.50, tension=0.1}{v2,v3}
  \fmf{fermion, right=0.25, tension=18}{v3,v4}
  \fmf{phantom, tension=20}{v2,r}
  \fmf{phantom, tension=20}{v3,r}
  \fmffreeze
  \fmf{fermion}{i3,v2}
  \fmf{fermion}{i4,v3}
  \fmf{fermion}{v3,o3}
  \fmf{fermion}{v2,o4}
  \fmfv{d.shape=circle,d.filled=full,d.size=2thick}{v1}
  \fmfv{d.shape=circle,d.filled=full,d.size=2thick}{v2}
  \fmfv{d.shape=circle,d.filled=full,d.size=2thick}{v3}
  \fmfv{d.shape=circle,d.filled=full,d.size=2thick}{v4}
  \end{fmfgraph}
  \end{fmffile}
\end{center}
}
      &
-3    \\
\hline
4 &
\parbox{50pt}{
\begin{center}
  \begin{fmffile}{Figures/diags/4thOrder1PI_RPA_2}
  \begin{fmfgraph}(30,80)
    \fmfkeep{4thOrder1PI_RPA_2}
    \fmfbottom{i1,i2,i3,i4}
    \fmftop{o1,o2,o3,o4}
    \fmfright{r1,r2,r3,r4}
    \fmf{fermion, tension=4}{i1,v1}
    \fmf{fermion}{v1,v3}
    \fmf{fermion, tension=4}{v3,o1}
    \fmffreeze
    \fmf{fermion, right=0.25}{v1,v2}
    \fmf{phantom, right=0.25, tension=4}{v2,v3}
    \fmf{phantom, right=0.25, tension=4}{v3,v2}
    \fmf{fermion, left=0.25, tension=10}{v2,r2}
    \fmf{phantom, left=0.25, tension=1}{v2,r4}
    \fmffreeze
    \fmf{fermion, right=0.35}{r2,v3}
    \fmf{fermion, left=0.75}{v3,r2}
    \fmf{fermion, right=0.25}{v2,v1}
    \fmf{fermion, left=0.25}{r2,v2}
    \fmfv{d.shape=circle,d.filled=full,d.size=2thick}{v1}
    \fmfv{d.shape=circle,d.filled=full,d.size=2thick}{v2}
    \fmfv{d.shape=circle,d.filled=full,d.size=2thick}{v3}
    \fmfv{d.shape=circle,d.filled=full,d.size=2thick}{r2}
  \end{fmfgraph}
  \end{fmffile}
\end{center}
}
&
3
&
\parbox{50pt}{
\begin{center}
  \begin{fmffile}{Figures/diags/4thOrder1PI_RPA_2_Cut}
  \begin{fmfgraph}(30,80)
    \fmfkeep{4thOrder1PI_RPA_2_Cut}
  \fmfstraight
  \fmfbottom{i1,i2,i3,i4}
  \fmftop{o1,o2,o3,o4}
  \fmfright{r}
  \fmf{fermion, tension=3}{i1,v1}
  \fmf{fermion}{i2,v1}
  \fmf{fermion}{i3,v2}
  \fmf{fermion}{i4,v2}
  \fmf{fermion, tension=2}{v1,v3}
  \fmf{fermion, tension=2}{v2,v3}
  \fmf{fermion, tension=1}{v3,o3}
  \fmf{fermion, tension=1}{v3,o4}
\fmffreeze
  \fmf{fermion, tension=2}{v4,o1}
  \fmf{fermion, tension=2}{v4,o2}
  \fmf{fermion, left=0.20}{v1,v4}
  \fmf{fermion, right=0.60}{v2,v4}
  \fmfv{d.shape=circle,d.filled=full,d.size=2thick}{v1}
  \fmfv{d.shape=circle,d.filled=full,d.size=2thick}{v2}
  \fmfv{d.shape=circle,d.filled=full,d.size=2thick}{v3}
  \fmfv{d.shape=circle,d.filled=full,d.size=2thick}{v4}
  \end{fmfgraph}
  \end{fmffile}
\end{center}
}
      &
-3    \\

\hline

\end{tabular}

\caption{
Examples of diagrams $\mathcal{G}^{(\mathrm{A},1)}_n$, with $p$ hole lines and one particle loop,
contributing to $G^{(\mathrm{A},1)}$ in the RPA approximation.
Associated cut diagrams $\mathcal{G}^{(0,1+p)}_n$ and their superficial degree of divergence
$D(\mathcal{G}^{(0,1+p))}_n)$ are given.}
\label{RPA_Cut_Diag}
\end{table}

\begin{table}
\centering

\begin{tabular}{|c|c|c|c|}

\hline
 $n$ \hspace{0.15cm}&
 $\mathcal{G}^{(\mathrm{A},1)}_n \in \mathcal{S}^{(\mathrm{A},1)}_{\text{RPA}}$ &
 \hspace{0.0cm} $p$ \hspace{0.25cm}&
 $\mathcal{G}^{(0,1+p)}_n \in \mathcal{S}^{(0,1+p)}_{\text{RPA}}$
\\

\hline
2 &
\parbox{50pt}{
\begin{center}
  \begin{fmffile}{Figures/diags/2ndOrder1PI_RPA_NoLoopP}
  \begin{fmfgraph}(30,80)
    \fmfkeep{2ndOrder1PI_RPA_NoLoopP}
  \fmfbottom{i}
  \fmftop{o}
  \fmf{fermion, tag=1}{v1,i}
  \fmf{fermion, tag=3}{v2,v1}
  \fmf{fermion, tag=5}{o,v2}
  \fmffreeze
  \fmf{fermion, right, tag=4}{v1,v2}
  \fmffreeze
  \fmf{fermion, right, tag=2}{v2,v1}
  \fmfv{d.shape=circle,d.filled=full,d.size=2thick}{v1}
  \fmfv{d.shape=circle,d.filled=full,d.size=2thick}{v2}
  \end{fmfgraph}
  \end{fmffile}
\end{center}
}
      &
2
      &
\parbox{50pt}{
\begin{center}
  \begin{fmffile}{Figures/diags/2ndOrder1PI_RPA_Cut_NoLoopP}
  \begin{fmfgraph}(30,80)
    \fmfkeep{2ndOrder1PI_RPA_Cut_NoLoopP}
    \fmfstraight
  \fmfbottom{i1,i2,i3,i4}
  \fmftop{o1,o2,o3,o4}
  \fmf{fermion}{i2,v1}
  \fmf{fermion}{i4,v1}
  \fmf{fermion}{v1,v2}
  \fmf{fermion}{v2,o1}
  \fmf{fermion}{v2,o3}
  \fmffreeze
  \fmf{fermion}{i1,v2}
  \fmf{fermion}{v1,o4}
  \fmfv{d.shape=circle,d.filled=full,d.size=2thick}{v1}
  \fmfv{d.shape=circle,d.filled=full,d.size=2thick}{v2}
  \end{fmfgraph}
  \end{fmffile}
\end{center}
}
 \\
\hline
3 &
\parbox{50pt}{
\begin{center}
  \begin{fmffile}{Figures/diags/3rdOrder1PI_RPA_1_NoLoopP}
  \begin{fmfgraph}(30,80)
    \fmfkeep{3rdOrder1PI_RPA_1_NoLoopP}
    \fmfbottom{i}
    \fmftop{o}
    \fmfright{r}
    \fmf{fermion, tension=4}{v1,i}
    \fmf{fermion}{v3,v1}
    \fmf{fermion, tension=4}{o,v3}
    \fmffreeze
    \fmf{fermion, right=0.25}{v1,v2}
    \fmf{fermion, right=0.25}{v2,v3}
    \fmf{phantom, tension=20}{v2,r}
    \fmffreeze
    \fmf{fermion, right=0.25}{v3,v2}
    \fmf{fermion, right=0.25}{v2,v1}
    \fmfv{d.shape=circle,d.filled=full,d.size=2thick}{v1}
    \fmfv{d.shape=circle,d.filled=full,d.size=2thick}{v2}
    \fmfv{d.shape=circle,d.filled=full,d.size=2thick}{v3}
  \end{fmfgraph}
  \end{fmffile}
\end{center}
}
&
3
&
\parbox{50pt}{
\begin{center}
  \begin{fmffile}{Figures/diags/3rdOrder1PI_RPA_1_Cut_NoLoopP}
  \begin{fmfgraph}(30,80)
    \fmfkeep{3rdOrder1PI_RPA_1_Cut_NoLoopP}
  \fmfstraight
  \fmfbottom{i1,i2,i3,i4,i5,i6}
  \fmftop{o1,o2,o3,o4,o5,o6}
  \fmfright{r}
  \fmf{fermion}{i3,v1}
  \fmf{fermion}{i5,v1}
  \fmf{fermion}{v1,v2,v3}
  \fmf{fermion}{v3,o2}
  \fmf{fermion}{v3,o4}
\fmffreeze
  \fmf{fermion}{i1,v3}
  \fmf{fermion}{i2,v2}
  \fmf{fermion}{v1,o6}
  \fmf{fermion}{v2,o5}
  \fmfv{d.shape=circle,d.filled=full,d.size=2thick}{v1}
  \fmfv{d.shape=circle,d.filled=full,d.size=2thick}{v2}
  \fmfv{d.shape=circle,d.filled=full,d.size=2thick}{v3}
  \end{fmfgraph}
  \end{fmffile}
\end{center}
}
   \\

\hline

\end{tabular}

\caption{
Examples of diagrams $\mathcal{G}^{(\mathrm{A},1)}_n$, with $p$ hole lines and no particle loop,
contributing to $G^{(\mathrm{A},1)}$ in the RPA approximation.
Associated cut diagrams $\mathcal{G}^{(0,1+p)}_n$ are given.}
\label{RPA_Cut_Diag_NoLoopP}
\end{table}

\begin{figure}
  \centering
\hspace{0.5cm}
\parbox{50pt}{
\begin{fmffile}{Figures/diags/RPA_divDiag}
\begin{fmfgraph*}(30,80)
  \fmfbottom{i1,i2}
  \fmftop{o1,o2}
  \fmf{fermion}{i1,v1}
  \fmf{fermion}{i2,v1}
  \fmf{fermion}{v2,o1}
  \fmf{fermion}{v2,o2}
\fmf{fermion, left=0.75}{v1,v2}
\fmf{fermion, right=0.75}{v1,v2}
\fmfv{d.shape=circle,d.filled=full,d.size=3thick,
      label=$C^R_0$,l.angle=-160,l.dist=5thick}{v1}
\fmfv{d.shape=circle,d.filled=full,d.size=3thick,
      label=$C^R_0$,l.angle=160,l.dist=5thick}{v2}
\end{fmfgraph*}
\end{fmffile}
}
\hspace{0.2cm}
+
\hspace{0.5cm}
\parbox{50pt}{
\begin{fmffile}{Figures/diags/RPA_divDiag_ct}
\begin{fmfgraph*}(30,80)
\fmfbottom{i1,i2}
\fmftop{o1,o2}
\fmf{fermion}{i1,v1}
\fmf{fermion}{i2,v1}
\fmf{fermion}{v1,o1}
\fmf{fermion}{v1,o2}
\fmfv{d.shape=circle,d.filled=hatched,d.size=5thick,
      label=$\delta C_0^{\text{RPA}}(\Lambda)$,l.angle=-0,l.dist=8thick}{v1}
\end{fmfgraph*}
\end{fmffile}
}
\caption{The only UV divergent diagram appearing in the RPA. Its associated counterterm is added.
The filled vertex represents the renormalized finite LEC $C^R_0$ while the hatched vertex represents
the counterterm $\delta C_0^{\text{RPA}}(\Lambda)$.}
\label{RPA_divDiag_ct}
\end{figure}
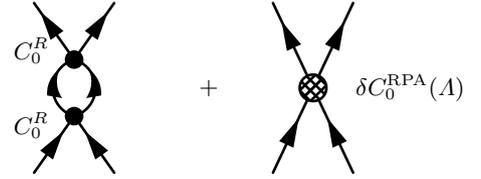

Applying the procedure designed in \refsec{subsec:GeneralProcedure} to 1PI time-ordered
diagrams, the set of diagrams with $p$ hole lines
belonging to $\mathcal{S}^{(\mathrm{A},1+p)}_{\text{RPA}}$ is explicitly
pictured at second, third and fourth orders in Tables~\ref{RPA_Cut_Diag} and~\ref{RPA_Cut_Diag_NoLoopP}.
The resulting diagrams displayed in \reftab{RPA_Cut_Diag_NoLoopP}
contain no loop and, thus, are free of any UV-divergence.
Those displayed in \reftab{RPA_Cut_Diag} read as one-loop diagrams
(made of $n$ internal lines) contributing to the $n$-body Green's function defined with respect to $\ket{0}$.
With the notations of \refsec{subsec:GeneralProcedure}, one has
\begin{subequations}
\begin{align}
  k &= 1 \ , \\
  p &= n-1 \ .
\end{align}
\end{subequations}
Following the BPHZ procedure for a diagram $\mathcal{G}^{(0,n)}_{n}~\in~\mathcal{S}^{(0,n)}_{\text{RPA}}$
is straightforward.
As $\mathcal{G}^{(0,n)}_{n}$ contains only one loop, any potential renormalization part $\gamma$ must contain
at least all $n$ internal lines building the loop so that
\begin{subequations}
\begin{align}
 L^\gamma &= 1 \ , \\
 I^\gamma &\ge n \ , \\
 n^\gamma &\le n \ ,
\end{align}
\end{subequations}
where $L^{\gamma}$ is the number of loops, $I^{\gamma}$ the number of internal lines and
$n^{\gamma}$ the number of vertices of $\gamma$.
Using the topological identity $L^{\gamma} = I^{\gamma} - n^{\gamma} + 1$ implies that
$I^{\gamma} = n^{\gamma}$ so that
\begin{subequations}
\begin{align}
  L^\gamma &= 1 \ , \\
  I^\gamma &= n \ , \\
  n^\gamma &= n \ .
\end{align}
\end{subequations}
Eventually, the only potential renormalization part of $\mathcal{G}^{(0,n)}_{n}$ is
$\mathcal{G}^{(0,n)}_{n}$ itself.
From \refeq{Simplified_SuperficialD}, the superficial degree of divergence of $\mathcal{G}^{(0,n)}_{n}$
for $n\ge 2$ reads as
\begin{equation}\label{RPA_D}
  D(\mathcal{G}^{(0,n)}_{n}) =  5 - 2 n \ .
\end{equation}
The only solution to $D(\mathcal{G}^{(0,n)}_{n}) \geq 0 $ is obtained for $n=2$.
Consequently, there is only one UV divergent diagram which is pictured in~\reffig{RPA_divDiag_ct} along with
the additional counterterm generated via the application of BPHZ.

The counterterm depends on the chosen renormalization scheme. Starting from the regularisation of \refeq{regMomSep}, the subtraction of the UV-divergence can be achieved by a zero-derivative contact counterterm $\delta C_0^{\text{RPA}}(\Lambda)$.
Matching the one-loop in-vacuum two-body Green's function to the experimental S-wave scattering length $a_0$ leads to
\begin{subequations}
\begin{align}
  C^R_0 &= \frac{4\pi}{m} a_0 \ , \label{RenoLEC}\\
  \delta C_0^{\text{RPA}}(\Lambda) &= \frac{4\pi}{m} \
    \frac{2}{\pi} \ \left( \int^{+\infty}_{0} \mathrm{d}q \ v^2_\Lambda(2q) \right)\ a_0^2
  \label{Ct_RPA} \ .
\end{align}
\end{subequations}

\begin{figure}
\centering
\parbox{50pt}{\begin{fmffile}{Figures/diags/Ct_1PI_1GF_RPA}
  \begin{fmfgraph}(30,80)
    \fmfkeep{Ct_1PI_1GF_RPA}
  \fmfbottom{i}
  \fmftop{o}
  \fmf{fermion}{i,v1}
  \fmf{fermion}{v1,o}
  \fmf{fermion, tension=1}{v1,v1}
  \fmfv{d.shape=circle,d.filled=hatched,d.size=5thick}{v1}
  \end{fmfgraph}
\end{fmffile}}
\hspace{-0.25cm}
+
\hspace{-0.4cm}
\parbox{50pt}{\begin{fmffile}{Figures/diags/1stOrder1PI_1GF_RPA}
  \begin{fmfgraph}(30,80)
    \fmfkeep{1stOrder1PI_1GF_RPA}
  \fmfbottom{i}
  \fmftop{o}
  \fmf{fermion}{i,v1}
  \fmf{fermion}{v1,o}
  \fmf{fermion, tension=1}{v1,v1}
  \fmfv{d.shape=circle,d.filled=full,d.size=2thick}{v1}
  \end{fmfgraph}
\end{fmffile}}
\hspace{-0.25cm}
+
\hspace{-0cm}
\parbox{50pt}{\begin{fmffile}{Figures/diags/2ndOrder1PI_1GF_RPA}
  \begin{fmfgraph}(30,80)
    \fmfkeep{2ndOrder1PI_1GF_RPA}
  \fmfbottom{i}
  \fmftop{o}
  \fmf{fermion, tag=1}{i,v1}
  \fmf{fermion, tag=3}{v1,v2}
  \fmf{fermion, tag=5}{v2,o}
  \fmffreeze
  \fmf{fermion, right, tag=4}{v1,v2}
  \fmffreeze
  \fmf{fermion, right, tag=2}{v2,v1}
  \fmfv{d.shape=circle,d.filled=full,d.size=2thick}{v1}
  \fmfv{d.shape=circle,d.filled=full,d.size=2thick}{v2}
  \end{fmfgraph}
\end{fmffile}}
\hspace{-0.65cm}
+
\hspace{-0.4cm}
\parbox{50pt}{\begin{fmffile}{Figures/diags/3rdOrder1PI_1GF_RPA_1}
  \begin{fmfgraph}(30,80)
    \fmfkeep{3rdOrder1PI_1GF_RPA_1}
  \fmfbottom{i}
  \fmftop{o}
  \fmfright{r}
  \fmf{fermion, tension=4}{i,v1}
  \fmf{fermion}{v1,v3}
  \fmf{fermion, tension=4}{v3,o}
  \fmffreeze
  \fmf{fermion, right=0.25}{v1,v2}
  \fmf{fermion, right=0.25}{v2,v3}
  \fmf{phantom, tension=20}{v2,r}
  \fmffreeze
  \fmf{fermion, right=0.25}{v3,v2}
  \fmf{fermion, right=0.25}{v2,v1}
  \fmfv{d.shape=circle,d.filled=full,d.size=2thick}{v1}
  \fmfv{d.shape=circle,d.filled=full,d.size=2thick}{v2}
  \fmfv{d.shape=circle,d.filled=full,d.size=2thick}{v3}
  \end{fmfgraph}
\end{fmffile}}
\hspace{-0.6cm}
+
\hspace{-0.4cm}
\parbox{50pt}{\begin{fmffile}{Figures/diags/4thOrder1PI_1GF_RPA}
  \begin{fmfgraph}(30,80)
    \fmfkeep{4thOrder1PI_1GF_RPA}
    \fmfbottom{i}
    \fmftop{o}
    \fmfright{r}
    \fmf{fermion, tension=5}{i,v1}
    \fmf{fermion}{v1,v4}
    \fmf{fermion, tension=5}{v4,o}
    \fmffreeze
    \fmf{fermion, right=0.25, tension=13}{v1,v2}
    \fmf{fermion, right=0.25, tension=0.1}{v2,v3}
    \fmf{fermion, right=0.25, tension=13}{v3,v4}
    \fmf{phantom, tension=30}{v2,r}
    \fmf{phantom, tension=30}{v3,r}
    \fmffreeze
    \fmf{fermion, right=0.25}{v4,v3}
    \fmf{fermion, right=0.25}{v3,v2}
    \fmf{fermion, right=0.25}{v2,v1}
    \fmfv{d.shape=circle,d.filled=full,d.size=2thick}{v1}
    \fmfv{d.shape=circle,d.filled=full,d.size=2thick}{v2}
    \fmfv{d.shape=circle,d.filled=full,d.size=2thick}{v3}
    \fmfv{d.shape=circle,d.filled=full,d.size=2thick}{v4}
  \end{fmfgraph}
\end{fmffile}}
\hspace{-0.6cm}
+
\dots

\caption{Time-unordered 1PI diagrams contributing to the one-body Green's function
in the RPA approximation with the additional
counterterm derived following the procedure in \refsec{subsec:GeneralProcedure}.}
\label{Fig_1GF_1PI_RPA}
\end{figure}
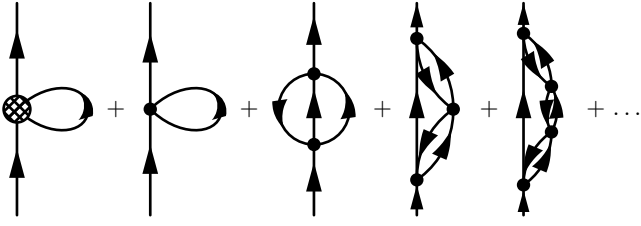

When computing the one-body Green's function within $\spi$EFT at LO in neutron matter at the RPA level, one concludes that
an additional diagram is required with the pure contact counterterm \eqref{Ct_RPA} to achieve renormalization.
The final set of RPA 1PI time-unordered diagrams contributing to the in-medium propagator are pictured in
\reffig{Fig_1GF_1PI_RPA}. The scheme is applicable to any A.


\subsection{Hartree-Fock partitioning}
\label{subsec:HFPart}

So far, MBPT has been formulated on the basis of choosing the kinetic energy as the unperturbed Hamiltonian. In the present section, a different partitioning of the Hamiltonian is considered to illustrate the flexibility of the procedure derived in \refsec{subsec:GeneralProcedure}.

Another partitioning of the Hamiltonian that plays an important role in nuclear physics is
\begin{subequations}\label{HF_PT}
\begin{align}
      H  &= H_0 + H_1 \ , \\
      H_0 &\equiv \sum_{\vec{p}\sigma}
                              e^{\text{HF}}_{\vec{p}\sigma}
                              a^\dagger_{\vec{p} \sigma}
                              a_{\vec{p} \sigma} \ , \\
      H_1 &\equiv
          \sum_{\vec{p}\sigma}
                              \left(\frac{p^2}{2m} - e^{\text{HF}}_{\vec{p}\sigma}\right)
                              a^\dagger_{\vec{p} \sigma}
                              a_{\vec{p} \sigma} \nonumber \\
          &\phantom{\equiv} +
          \frac{1}{2!} \sum_{\sigma_1 \sigma_2}
          \sum_{\substack{\vec{p}_1 \vec{p}_2 \\ \vec{p}_1^\exprime \vec{p}_2^\exprime}} \
          (2\pi)^3 \delta(\vec{p}_1^\exprime + \vec{p}_2^\exprime - \vec{p}_1 - \vec{p}_2) \ C_0
          \nonumber \\
          &\phantom{\equiv + \frac{1}{2!} \sum \sum \quad }
          a^\dagger_{\vec{p}_1^\exprime \sigma_1}
          a^\dagger_{\vec{p}_2^\exprime \sigma_2}
          a_{\vec{p}_1 \sigma_1}
          a_{\vec{p}_2 \sigma_2} \ ,
\end{align}
\end{subequations}
where $e^{\text{HF}}_{\vec{p}\sigma}$ denotes Hartree-Fock (HF) single-particle energies.
Let us consider homogeneous neutron matter of density $\rho$ and use the reference Slater determinant
\begin{equation}\label{FermiSeakF}
| \Phi^\rho_0 \rangle \equiv \prod_{\sigma , p < k_F} a^\dagger_{\vec{p}\sigma} \ket{0} \ ,
\end{equation}
with $k_F \equiv \left(3\pi^2 \rho\right)^{1/3}$ the associated Fermi momentum.
The HF in-medium propagator reads as
\begin{equation}
  G^{(\rho,1)\text{HF}}_{\vec{p}\sigma}(\omega) =
   \frac{\theta\left( p - k_F \right)}{\omega - e^{\text{HF}}_{\vec{p}\sigma} + i\eta}
            + \frac{\theta\left( k_F - p \right)}{\omega - e^{\text{HF}}_{\vec{p}\sigma} - i\eta} \ ,
\end{equation}
with HF self-energy $\Sigma^{(\rho)\text{HF}}_{\vec{p}\sigma}$ being defined as
\begin{equation}
  \Sigma^{(\rho)\text{HF}}_{\vec{p}\sigma} \equiv \frac{p^2}{2m} - e^{\text{HF}}_{\vec{p}\sigma} \ ,
\end{equation}
which is energy-independent.
In the case of a simple contact interaction, the HF self-energy is even independent of the momentum
and reads as
\begin{equation}\label{HF_selfEnergy}
  \Sigma^{(\rho)\text{HF}}_{\vec{p}\sigma}
    \equiv \Sigma^{(\rho)\text{HF}} = \rho \left(1-\frac{1}{2}\right) C_0 \ .
\end{equation}
Consequently, the unperturbed particle propagator reads as
\begin{equation}
  G^{(\rho,1)\text{HF}+}_{\vec{p}\sigma}(\omega) =
   \frac{\theta\left( p - k_F \right)}{\omega - (\frac{p^2}{2m} + \rho \left(1-\frac{1}{2}\right) C_0) + i\eta}
\end{equation}
and its asymptotic coefficients $\alpha^{\text{HF}}\left(S\right)$ verify
\begin{equation}
  \alpha^{\text{HF}}\left(S\right) =
   \begin{cases}
     -1 &\text{ if } S = \{ \vec{e}_\omega\} \\
     -2 &\text{ if } S = \{ \vec{L} \} \text{ with } \vec{L} \notin \{\vec{e}_\omega\} \\
     -2 &\text{ if } \dim{S} \geq 2
   \end{cases}
\ .
\end{equation}
Eventually, as $\alpha^{\text{HF}}\left(S\right) = \alpha^0\left(S\right)$,
the procedure described in \refsec{subsec:GeneralProcedure} can be applied for the partitioning \eqref{HF_PT}.

As one is interested in adding corrections to HF, the Hamiltonian is regularised and contains counterterms, \textit{i.e.}\begin{subequations}\label{RegHF_PT}
\begin{align}
      H  &= H_0 + H_1 \ , \\
      H_0 &\equiv \sum_{\vec{p}\sigma}
                              e^{\text{HF}}_{\vec{p}\sigma}(\Lambda) \
                              a^\dagger_{\vec{p} \sigma}
                              a_{\vec{p} \sigma} \ , \\
      H_1 &\equiv
          \sum_{\vec{p}\sigma}
                              \left(\frac{p^2}{2m} - e^{\text{HF}}_{\vec{p}\sigma}(\Lambda)\right)
                              a^\dagger_{\vec{p} \sigma}
                              a_{\vec{p} \sigma} \nonumber \\
          &\phantom{\equiv} +
          \frac{1}{2!} \sum_{\sigma_1 \sigma_2}
          \sum_{\substack{\vec{p}_1 \vec{p}_2 \\ \vec{p}_1^\exprime \vec{p}_2^\exprime}} \
          (2\pi)^3 \delta(\vec{p}_1^\exprime + \vec{p}_2^\exprime - \vec{p}_1 - \vec{p}_2) \ C_0(\Lambda)
          \nonumber \\
          &\phantom{\equiv + \frac{1}{2!} \sum \sum \  }
          \times v_\Lambda(\vec{p}_1 - \vec{p}_1^\exprime) \
          a^\dagger_{\vec{p}_1^\exprime \sigma_1}
          a^\dagger_{\vec{p}_2^\exprime \sigma_2}
          a_{\vec{p}_1 \sigma_1}
          a_{\vec{p}_2 \sigma_2} \ .
\end{align}
\end{subequations}
Note that in \refeqs{RegHF_PT} the chosen regularisation is different from \refeq{RegH}.
As the regularisation will not impact many-body calculations, it should be chosen based on convenience.
In this case, the regularised potential only depends on momentum transfer $\vec{p}_1^\exprime - \vec{p}_1$
and the HF self-energy reads simply as
\begin{equation}\label{RegHF_selfEnergy}
  \Sigma^{(\rho)\text{HF}}_{\vec{p}\sigma}(\Lambda)
    = C_0(\Lambda)\left(\rho  - \int_{\abs{\vec{p}^\exprime} < k_F} \frac{\mathrm{d}^3\vec{p}^\exprime}{(2\pi)^3} \
                        v_\Lambda(\vec{p} - \vec{p}^\exprime) \right) \ .
\end{equation}
The self-energy \eqref{RegHF_selfEnergy} is now momentum-dependent.
In general, as long as
\begin{subequations}\label{sufficientConditions}
\begin{align}
  \Sigma^{(\rho)\text{approx}}_{\vec{p}\sigma}(\omega ; \Lambda)
    &\underset{p \to +\infty }{=} O\left( p^2 \right) \ , \\
  \Sigma^{(\rho)\text{approx}}_{\vec{p}\sigma}(\omega ; \Lambda)
    &\underset{\omega \to +\infty }{=} O\left( \omega \right) \ ,
\end{align}
\end{subequations}
the renormalization procedure of \refsec{subsec:GeneralProcedure} can be applied.
In the case of Eq.~\eqref{RegHF_selfEnergy}, using \refeq{limitUVsupp} is sufficient for
\refeqs{sufficientConditions} to hold. For derivative contact interactions
it becomes necessary to take $v_\Lambda$ decreasing sufficiently fast, \textit{e.g.}\ exponentially.

\section{Conclusions}
\label{sec:Ccl}

Starting from the derivation of a Hamiltonian $H$ describing the interaction between nucleons within an EFT approach (namely $\spi$EFT), the power-counting rules proposed to compute observables at LO require to exactly solve the $\mathrm{A}$-body Schr\"{o}dinger equation for the truncated Hamiltonian $H^{\text{LO}}$ in such a way that renormalization is ensured. However, exact calculations remain intractable in large $\mathrm{A}$-body sectors ($\mathrm{A} \gg 10$), which may compromise the renormalization invariance of computed observables. In order to overcome this tension in the context of many-body methods that can be formulated in terms of time-ordered diagrams, the idea pursued in this article is to design a method to renormalize workable truncations and check whether or not this renormalization procedure is consistent with the one traditionally employed in $\spi$EFT via the exact calculation of few-body systems. The answer to this question depends on the truncation scheme employed.

In this article, a method is thus designed to identify divergences for any given set of many-body diagrams generated through a perturbative expansion of the $k$-body Green's function around an $\mathrm{A}$-body Slater determinant reference state. This set can be strictly perturbative or eventually correspond to the resummation of an infinite (subset) of diagrams, \textit{e.g.}\ summing particle-particle or particle-hole ladders. The method involves a so-called 'cutting' procedure allowing one to relate the ultraviolet divergences of the in-medium $k$-body Green's function to those displayed by a set of \emph{in-vacuum} $(k+p)$-body Green's functions. Applying the BPHZ procedure to the diagrams making up the in-vacuum Green's functions, counterterms necessary to renormalize the original in-medium $k$-body Green's function are identified in a systematic fashion. This procedure delivers the desired property that $k$-body counterterms are independent of the $\mathrm{A}$-body sector ($A\geq k$) one starts from. Eventually, the present development is similar to what has been done in QFT at finite temperature~\cite{Collins1984,LeBellac1996}.

This work only constitutes a first step forward and critical extensions remain to be carried out.
First, one must go from perturbative to intrinsically non-perturbative methods based on in-medium diagrams such as CC, SCGF or IM-SRG. Regarding SCGF, let us mention the important work showing possible additional counterterms compared to the naive application of BPHZ to diagrams formulated in terms of fully-dressed propagators~\cite{VanHees2002a,Blaizot2004,Berges2005}. Furthermore, additional steps are needed to extend the present developments to more general partitioning.
Second, of importance are partitionings based on unperturbed Hamiltonians breaking exact symmetries of the Hamiltonian. The most trivial case consists of breaking translational invariance such that unperturbed propagators are no longer diagonal in momentum space, \textit{e.g.}\ using a harmonic oscillator Hamiltonian. The use of Weinberg's asymptotic theorem must be extended in such a case. Less trivial are partitionings breaking $U(1)$ symmetry that are employed to tackle the superfluid character of nuclear matter and open-shell nuclei, \textit{e.g.}\ MBPT~\cite{Duguet:2015yle,Tichai:2018mll,Arthuis:2018yoo}, CC~\cite{Si15} or SCGF~\cite{So11,Soma14b} using a Bogoliubov vacuum as reference state. The associated diagrammatic relies on the use of anomalous propagators in addition to normal propagators.
As a result, the analysis of ultraviolet divergencies is fundamentally different so that diagrams contributing to the mean-field (\textit{i.e.}\ Hartree, Fock and Bogoliubov) already require counterterms~\cite{Gorkov1961,Papenbrock1999} contrary to those (\textit{i.e.}\ Hartree and Fock) at play in the perturbation theory considered here.

\section*{Acknowledgements}

The authors thank U. van Kolck, B. Long, M. Pav\'on Valderrama and C. Wellenhofer for insightful discussions.


\bibliographystyle{apsrev4-1}
\bibliography{EPJA_Bib}

\end{document}